\documentclass[14pt]{article}
\pdfoutput=1
\usepackage{amsmath,amssymb,amsfonts,amsthm,bm,bbm,cancel,wasysym}
\usepackage{epsfig,graphics,graphicx,epstopdf}
\usepackage{array,booktabs,colortbl,colordvi,multirow}
\usepackage{colordvi,color,xcolor}
\usepackage{hyperref}
\usepackage{rotating}

\usepackage{verbatim}
\usepackage{cite}
\usepackage{subfig}
\usepackage{setspace}
\usepackage{url}
\usepackage[percent]{overpic}
\usepackage{slashed}
\usepackage{authblk}
\usepackage{xspace}
\usepackage{fullpage}
\usepackage{hyperref}

\def\ie{{\it i.e.}}
\def\eg{{\it e.g.}}
\def\etc{{\it etc}}

\def\to{\rightarrow}

\newskip\zatskip \zatskip=0pt plus0pt minus0pt
\def\matth{\mathsurround=0pt}
\def\lsim{\mathrel{\mathpalette\atversim<}}
\def\gsim{\mathrel{\mathpalette\atversim>}}
\def\atversim#1#2{\lower0.7ex\vbox{\baselineskip\zatskip\lineskip\zatskip
  \lineskiplimit 0pt\ialign{$\matth#1\hfil##\hfil$\crcr#2\crcr\sim\crcr}}}

\begin{document}


\begin{flushright}
SLAC-PUB-250226\\
\today
\end{flushright}
\vspace*{5mm}

\renewcommand{\thefootnote}{\fnsymbol{footnote}}
\setcounter{footnote}{1}

\begin{center}

{\Large {\bf Portal Matter Models of Kinetic Mixing with Two Light Dark Gauge Bosons}}\\

\vspace*{0.75cm}

{\bf Thomas G. Rizzo}~\footnote{rizzo@slac.stanford.edu}

\vspace{0.5cm}

{SLAC National Accelerator Laboratory}\\ 
{2575 Sand Hill Rd., Menlo Park, CA, 94025 USA}

\end{center}
\vspace*{.5cm}


\begin{abstract}
\noindent  

The kinetic mixing (KM) portal mandates the existence of at least one new gauge boson, the dark photon (DP) based on the group $U(1)_D$, which mixes with the Standard Model 
(SM) photon via loops of other new heavy particles carrying both SM and dark charges called portal matter (PM).  Arguments exist based on the RGE running of the $U(1)_D$ gauge coupling 
suggesting that at higher scales $U(1)_D$ becomes part of a more complex non-Abelian group, a simple example being just the SM-like $G_D=SU(2)_I \times U(1)_{Y_I}$.  In our past 
analyses it was always assumed that $G_D$ broke in a SM-like manner directly to the DP's $U(1)_D$ which then subsequently broke at low energies $\lsim 1$ GeV. However, this need not 
be the case and $G_D$ can instead break to $U(1)_{T_{3I}}\times U(1)_{Y_I}$, with $T_{3I}$ being the diagonal generator of $SU(2)_I$, now producing two light gauge bosons which obtain 
masses at the $\lsim 1$ GeV scale. In this paper we will explore the phenomenology of a very simple realization of this kind of alternative setup employing non-abelian KM and having a minimal, 
lepton-like PM sector, demonstrating its distinctive nature in comparison to the previously examined symmetry breaking path. The effects of interference between these gauge bosons on 
thermal dark matter annihilation, the production of new heavy gauge, Higgs and PM states at colliders, as well as the corresponding signatures for the light dark gauge bosons are examined.

\end{abstract}

\vspace{0.5cm}
\renewcommand{\thefootnote}{\arabic{footnote}}
\setcounter{footnote}{0}
\thispagestyle{empty}
\vfill
\newpage
\setcounter{page}{1}



\section{Introduction and Overview}

The Standard Model (SM), though very successful at describing a huge array of experimental results, fails to address some of our basic questions and observations. Perhaps the greatest challenge 
is the nature of dark matter (DM): what is it, why is it so common, and how does it interact with us? To address these and related mysteries necessarily leads us into vast realms outside of the 
SM.  Currently, the only established interaction of DM with the SM is via gravity, which is the manner in which DM first let us know of its existence via astrophysical and cosmological measurements. 
Under most circumstances, assuming that DM does not dominantly consist of primordial black holes\cite{Carr:2020xqk}, it is very likely that for DM to achieve its observed relic 
abundance\cite{Planck:2018vyg}, additional, yet to be observed, interaction(s) of some kind are necessary between it and ordinary matter. 

The possible scenarios for just what the DM may be go back many years in the history of particle physics and were traditionally associated with the solutions to other problems such CP-conservation 
in QCD and the gauge hierarchy problem, \ie, the QCD axion\cite{Kawasaki:2013ae,Graham:2015ouw,Irastorza:2018dyq} and thermal WIMPS, as appearing naturally in R-parity  
conserving supersymmetry, lying in the few GeV to $\sim 100$ TeV mass range\cite{Arcadi:2017kky,Roszkowski:2017nbc,Arcadi:2024ukq}. Ever widening searches for such states have gone on 
for many years, pushing deeper into model parameter spaces but have so far come up 
empty\cite{LHC,Aprile:2018dbl,Fermi-LAT:2016uux,Amole:2019fdf,LZ:2022ufs,PandaX:2024qfu,SuperCDMS:2024yiv,Aprile:2024xqi}. While such searches will continue for some time, these null 
results have prompted a host of new DM scenarios wherein candidates exist over wide regimes in both masses and 
couplings\cite{Alexander:2016aln,Battaglieri:2017aum,Bertone:2018krk,Cooley:2022ufh,Boveia:2022syt,Schuster:2021mlr,Cirelli:2024ssz}, presenting us will significant experimental challenges if 
we wish to explore a significant part of this large domain. However, for much of this parameter space, the necessary interactions between us and DM can be described by a relatively small set of 
models, termed `portals', which are essentially effective field theories (EFTs) of various classes that may or may not also be renormalizable. In all cases, in addition to the DM and SM fields, the 
existence of other new particles are required which act as mediators for this new interaction\cite{Lanfranchi:2020crw} but whose nature depends upon the details of a particular setup.
 
 One of the simplest and attractive models of this kind extends the idea of the traditional thermal WIMP setup\cite{Steigman:2012nb,Steigman:2015hda,Saikawa:2020swg} down into the lower 
 DM mass range of 
 a few MeV up to several GeV, something which is not possible if only the familiar SM forces exist due to the Lee-Weinberg bound\cite{Lee:1977ua,Kolb:1985nn}. In the renormalizable kinetic 
 mixing (KM) - vector portal model\cite{KM,vectorportal,Gherghetta:2019coi}, a new `dark' gauge interaction is introduced based on the group $U(1)_D$, whose (generally massive) gauge boson, 
 the dark photon (DP), $V$\cite{Fabbrichesi:2020wbt,Graham:2021ggy,Barducci:2021egn}, couples to a conserved dark charge, $Q_D$, that is carried by the DM but not by the SM fields.  The 
 generation of the DP mass via the breaking of $U(1)_D$ is assumed to occur via the vevs of one or more dark Higgs (DH) fields\cite{Li:2024wqj} as in the SM; these same vevs may also contribute 
 to the mass of the DM itself which,  depending upon the details of the setup,  is assumed to lie in the general proximity of that of the DP $\lsim 1$ GeV. The interaction between the SM with the dark 
 sector fields is then induced (at low energy scales) by the KM of the DP and the SM photon, $A$, at loop-level via vacuum polarization-like diagrams. For such diagrams to occur at, \eg, the 1-loop 
 level, other new fields, presumably heavy, must also be present which carry both SM as well as $U(1)_D$ dark charges so that they can couple to both $V$ and $A$. In order for such new states 
 to have avoided detection as well as to satisfy Higgs coupling, unitarity and precision electroweak constraints they must be either complex scalars and/or vector-life fermions 
 (VLFs)\cite{CarcamoHernandez:2023wzf,CMS:2024bni,Alves:2023ufm,Banerjee:2024zvg,Guedes:2021oqx,Adhikary:2024esf,Benbrik:2024fku,Albergaria:2024pji}. 
We have referred to such exotic particles as Portal Matter (PM) and they have been the subject of much recent attention by ourselves and other authors\cite{Rizzo:2018vlb,Rueter:2019wdf,Kim:2019oyh,Rueter:2020qhf,Wojcik:2020wgm,Rizzo:2021lob,Rizzo:2022qan,Wojcik:2022rtk,Rizzo:2022jti,Rizzo:2022lpm,Wojcik:2022woa,Carvunis:2022yur,Verma:2022nyd,Rizzo:2023qbj,Wojcik:2023ggt,Rizzo:2023kvy,Rizzo:2023djp,Rizzo:2024bhn,Ardu:2024bxg,Rizzo:2024kzu}. 

If the masses and couplings of these PM fields were known within a given setup, these 1-loop diagrams can be explicitly calculated and the strength of this KM, described by a dimensionless parameter 
$\epsilon$, can be determined, \ie,  
\begin{equation}
\epsilon =\frac{g_D e}{12\pi^2} \sum_i ~(\eta_i N_{c_i}Q_{em_i}Q_{D_i})~ ln \frac{m^2_i}{\mu^2}\,.
\end{equation}
where $g_D$ is the $U(1)_D$ gauge coupling (and so we can also by analogy define $\alpha_D=g_D^2/4\pi$), $m_i(Q_{em_i},Q_{D_i}, N_{c_i})$ are the mass (electric charge, dark 
charge, number of colors) of the $i^{th}$ PM field, while $\eta_i=1(1/4)$ if the PM is a Dirac VLF (complex scalar).  Note that $\epsilon$ is not generally a finite quantity as it explicitly depends upon the 
renormalization scale $\mu$;  {\it however} it is possible to have a more UV-complete arrangement wherein group theoretical constraints will render the sum inside the parenthesis above vanishing as 
in the setup below, \ie, 
\begin{equation}
\sum_i ~(\eta_i N_{c_i} Q_{em_i}Q_{D_i})=0\,,
\end{equation}
making $\epsilon$ both finite and calculable in such a model. Here we see that if $g_D \sim e$ and the PM fields are relatively close in mass we might expect that $\epsilon \sim 10^{-(3-4)}$ which is 
in the range needed to obtain the observed relic density while also being able to satisfy the so far null experimental search constraints for the DP. 
We also note that in such more complete setups, wherein PM and some of SM fields may share a representation of an enlarged gauge 
group, other loops graphs involving PM fields, together with the heavy gauge fields of this hypothetical group, can exist which generate additional 1-loop interactions between the SM and DM.

As already mentioned, models in this class must satisfy numerous constraints arising from a suite of different observations and experiments including those from direct and indirect detection as well  
as accelerator searches and also those from both astrophysics and cosmology. While we know that to obtain the observed abundance of DM the velocity-weighted annihilation cross section into SM 
particles at freeze-out must roughly satisfy the generic requirement\cite{Steigman:2012nb,Steigman:2015hda,Saikawa:2020swg}), $\sigma v_{rel}\sim 3\times 10^{-26}~$cm$^3$ sec$^{-1}$. We 
also know for DM this mass range, $\lsim 1$ GeV, that during the time of the CMB this cross section must be suppressed by several orders of 
magnitude\cite{Planck:2018vyg,Slatyer:2015jla,Liu:2016cnk,Leane:2018kjk,Wang:2025tdx} to avoid dumping additional 
electromagnetic energy into the thermal bath, and similarly so today\cite{Koechler:2023ual,DelaTorreLuque:2023cef,Wang:2025jhy}. This implies that $\sigma v_{rel}$ must have a significant 
temperature ($T$) dependence - something that powerfully excludes certain types of DM and interactions with the SM fields in this mass range. A general class of models is immediately 
impacted: those that involve a 
standard $s$-wave annihilation process as would be the case for Dirac fermion DM with a vector mediator, \eg, a DP, so that we must entertain other possibilities. For example, if this process were 
instead ($i$) $p-$wave so that there is a $v^2 \sim T$ suppression at later times, as may be realized in the case of complex scalar or Majorana fermionic 
DM{\footnote {See, however\cite{Belanger:2024bro}}} we may be able to avoid such constraints. Another possibility is that DM annihilation to SM final states may take place $(ii)$ through the 
co-annihilation mechanism, as would be the realized by pseudo-Dirac DM with a sizable mass splitting also leading to an inelastic DM setup; this could be induced by, \eg, a $Q_D=2$ dark Higgs 
vev. In such a case the DM annihilation rate is exponentially Boltzmann-suppressed, in a manner controlled by this mass splitting, due to the much lower temperatures at sufficiently 
later times\cite{Brahma:2023psr,Balan:2024cmq,Garcia:2024uwf,Mohlabeng:2024itu}.  As is well-known, if the mass splitting is sufficiently large in the inelastic DM setup, constraints arising from both 
direct and indirect detection searches can also be rather easily avoided.

It is interesting to note that, depending upon the details of the low energy field content, it has been observed that the RGE running of the $U(1)_D$ coupling (expressed through $\alpha_D$) into 
the UV may lead to the eventual loss of perturbativity at a scale not too far above the weak scale\cite{Davoudiasl:2015hxa,Reilly:2023frg,Rizzo:2022qan,Rizzo:2022lpm}. Clearly, this will be 
especially true of $\alpha_D$ starts out at largish values (above $\sim 0.2$, say) already at the DP mass ($\sim 100$ MeV or so) scale.  The obvious solution is that $U(1)_D$ must be 
embedded into some larger, non-abelian gauge group, $G_D$, whose beta-function will have the opposite sign, before this scale is reached. One may easily imagine that the PM mass scale and that 
associated with the breaking of $G_D$ may be intimately related. This possibility plus the investigation of how the SM and $G_D$ might be unified into a more UV-complete setup has been the 
subject of much of our recent work\cite{Rizzo:2018vlb,Rueter:2019wdf,Rueter:2020qhf,Wojcik:2020wgm,Rizzo:2021lob,Rizzo:2022qan,Rizzo:2022jti,Rizzo:2022lpm,Rizzo:2023qbj,Rizzo:2023kvy,Rizzo:2023djp,
Rizzo:2024bhn,Rizzo:2024kzu}. In some simple versions of these setups, it was assumed that $G_D$ was `SM-like' in that $G_D=SU(2)_I\times U(1)_{Y_I}$\cite{Bauer:2022nwt}, where here the 
label `I' refers to the almost invisible interactions taking place in the dark sector, with $U(1)_D$ embedded into $G_D$ exactly like $U(1)_{em}$ is in the case of the SM - except for it also being 
broken at small, $\sim 1$ GeV, mass scales{\footnote {This possibility was originally motivated by our earlier work on $E_6$-type gauge models\cite{Hewett:1988xc}.}} Further paralleling the SM, 
it was assumed that $G_D$ broke {\it directly} to only $U(1)_D$ in a single step but at the $\sim 10$ TeV scale so the heavy gauge bosons analogous to the SM $W,Z$ (here denoted as $Z_I,W_I$) 
would obtain comparable large masses as part of this process, leading to some interesting phenomenological possibilities at high energy colliders and elsewhere..

In this paper, we will explore an alternative possibility by giving up this `SM-like' bias in the manner in which $G_D$ breaks thus leading to a somewhat different set of phenomenological predictions. 
Here it will be assumed that at the $\sim 10$ TeV scale {\it only} $SU(2)_I$ is broken by a real scalar triplet with $Y_I=0$, $\Sigma$, whose $T_{3I}=0$ component gets a large vev, $v_T$, 
generating a mass $g_Iv_T$, for the $W_I^{\pm} \sim W_{1I}\pm iW_{2I}$ gauge bosons (analogs of the SM $W^\pm$) but leaving the $W_{3I}$ field massless.  Hence, below this scale, 
$U(1)_{T_{3I}}\times U(1)_{Y_I}$ will remain unbroken down to $\lsim 1$ GeV. The VLF PM chosen for this setup is likely the simplest possibility that one can imagine: an $SU(2)_I$ doublet, 
$SU(2)_L$ singlet, with $Y_I=0$ but having a non-zero SM hypercharge, $Y/2=-1$, in analogy with the right-handed charged 
leptons. The two $Q_{em}=-1$ states in this doublet, $E_{1,2}$ having $T_{3I}=\pm 1/2$, are initially degenerate in mass (by symmetry arguments), but this degeneracy is then broken by same large 
vev that breaks $SU(2)_I$ down to $U(1)_{T_{3I}}$. Once this degeneracy is broken, non-abelian KM at the 1-loop level between the SM hypercharge gauge boson, $B$, and the still massless 
$W_{3I}$ is generated; the value of $\epsilon$ that results from this is found to be finite and is a function of the ratio of the $E_{1,2}$ masses along the lines discussed above. A similar setup 
is also possible by employing instead color-triplet, $SU(2)_L$-singlet VL quarks with similar results. 
Subsequent to the usual SM symmetry breaking by the Higgs doublet at $\sim 246$ GeV, this KM (to leading order in the ratio of the relevant vevs) will manifest as 
occurring between the SM photon and the $W_{3I}$; once $U(1)_{T_{3I}}\times U(1)_{Y_I}$ is completely broken near $\sim 1$ GeV, the photon will appear to mix with both of these light gauge 
boson mass eigenstates, here denoted as $Z_{1,2}$, which are linear combinations of $W_{3I}$ and $B_I$. This final step of symmetry breaking at the low scale requires the introduction of a pair 
of dark Higgs fields with several possible choices available and which lead to somewhat different phenomenologies. We will argue the merits of a specific choice of dark Higgs that leads to 
pseudo-Dirac DM, so that the advantages of the inelastic DM setup can be realized, and that also generates a mixing between the lighter of the $E_i$ and, \eg, the SM $e_R$ to allow for this PM 
field to decay sufficiently rapidly. Since there are now two light 'dark' gauge bosons which will both couple to the SM via KM (instead of just the usual DP) interference effects will be 
shown to be of some relevance to the relic density calculation, especially so if they have comparable couplings and are not too different in mass, \ie, $M_{Z_2}/M_{Z_1} \lsim 2-2.5$. In our case, this 
will open up previously disfavored parameter space regions satisfying this observational constraint. We will also find that, in comparison to the previously examined `SM-like' $G_D$ breaking 
scenario, the minimal version of this type of setup consequently has a somewhat more limited high energy collider footprint. 

The outline of this paper is as follows:  After this Introduction, in Section 2 we present a detailed description of this alternative model setup following along with the effects of both the non-abelian KM 
and the various steps of gauge symmetry breaking down to the sub-GeV scale leading to there being two light dark gauge bosons, $Z_i$. For this specific setup, an $SU(2)_L$ isodoublet  of SM 
charged singlet VLF is chosen as the PM fields although we note that similar results are obtained with other possible choices.  Here we will argue that a specific minimal choice of the 
two Higgs fields that are necessary in this final step of symmetry can be made based on other phenomenological requirements that need to be satisfied: allowing the PM fields to decay and the 
generate pseudo-Dirac DM. This choice of Higgs fields then essentially fixes the relative (vector) couplings of the $Z_i$ once their $O(1)$ mass ratio is determined; we then explore the 
implications of this through a set of representative benchmark models. In Section 3, we study the impact of having these two light gauge bosons contribute to the the annihilation of DM into SM 
final states as part of the relic density calculation in the case where the DM is assumed to be pseudo-Dirac/inelastic in nature with a significant mass splitting; this choice is made in order to easily 
satisfy several existing observational and experimental constraints. We will show that when the $Z_i$ are relatively close in mass, constructive interference between the exchanges opens up a 
newly favorable region of parameter space, $m_{DM}/M_{Z_1} \sim 0.5-0.9$, which had been previously disfavored. In Section 4, we will explore some of the collider, \ie, LHC,  FCC-hh and 
$e^+e^-$ signatures that arise in the current setup. The new heavy states include the SM isosinglet, lepton-like PM fermions, $E_i$, as well as the $W_I^\pm$ that result from 
$SU(2)_I \to U(1)_{T_{3I}}$ breaking. While the $E_i$ are pair-produced as usual via SM electroweak interactions and decay into $e$ + MET, the $W_I$'s are not 
so easily produced here as in the previously examined setup as there this relied upon ($i$) the PM sharing an $SU(2)_I$ representation with some SM field and/or ($ii$) the existence of a heavy 
$Z_I$ state which is coupled to the SM as well as to $W_I$ pairs. Neither of these conditions hold in the current model. Here, we show that $W_I$ production is instead $\epsilon^2$ suppressed and 
so likely to be  inaccessible at a hadron collider unless the $W_I$is rather light and/or large center of mass energies are available.  
Other possible states in the dark scalar sector, most notably the heavy real scalar, $\Sigma^0$, which is left over from this same $SU(2)_I$ breaking, may also be made visible via its mixing with 
the SM Higgs. This same mixing is then shown to generate rare Higgs decays, such as $H(125)\to \gamma +Z_i$, but with branching fractions that are expected to be below $\sim 10^{-8}$ and are 
thus unfortunately invisible, far below the SM background rate. We will also show that the production of the $Z_i$ in $e^+e^-$ via PM exchange is difficult to observe without precise understanding 
of the SM background. Finally, a discussion and our conclusions can be found in Section 5.


\section{A Model with Two Light Dark Gauge Bosons: The Basic Setup}

In our past discussions where we have considered a dark sector gauge group of the form $G_D=SU(2)_I \times U(1)_{Y_I}=2_I1_{Y_I}$ (with the corresponding gauge bosons being 
$W_{i(=1-3)I},B_I$ having the corresponding couplings $g_I,g_{Y_I}$, respectively), we have always assumed that the symmetry breaking took place in a rather familiar, `SM-like' manner, 
\ie,  $2_I1_{Y_I}\to 1_D$. This was assumed to occur at the scale of a $\sim $ few TeV or more via a dark Higgs, $2_I$ doublet vev, with the $1_D$ remaining unbroken and whose gauge boson 
is identified with the dark photon (DP), $V$. This $U(1)_D$ group was subsequently broken near the $\sim 1$ GeV scale via a vev of another dark Higgs that carries a non-zero dark charge, 
$Q_D\neq 0$. This DP then undergoes KM via PM loops with one or both of the neutral SM gauge fields, which at energies far below the weak scale, then emerges as an effective kinetic mixing 
with just the SM photon. Of course, even if we assume that the dark gauge group, $G_D$, remains as above, it is of course always possible that this symmetry may be broken in an alternative, 
non-SM-like manner leading to a somewhat different phenomenology that we will now investigate below in a simple setup,  removing this SM bias. 

We now imagine that at the high scale, $\sim$ a few TeV or more, $G_D$ is broken by the $T_{3I}=0$ vev, $v_T$, of an $SU(2)_I$ triplet, $\Sigma$, carrying $Y_I=0$. This vev generates 
the masses, $g_Iv_T$, for the non-hermitian gauge bosons $W_I^\pm${~\footnote {Here, the $\pm$ label refers not to the dark or electric charge but to the corresponding dark isospin raising 
and lowering 
operators, $T_I^\pm$, to which $W_I^\pm$ couple.}} but clearly leaves both of the two remaining $W_{3I}$ and $B_I$ gauge fields massless, \ie, the group $U(1)_{T_{3I}}\times U(1)_{Y_I}$ 
remains unbroken. The fields in $\Sigma$ with $T_{3I}=\pm 1$, in the large vev limit, essentially 
then become the Goldstone bosons for these now massive $W_I^\pm$ and no longer remain in the physical spectrum. Simultaneously with this, we also imagine the existence of an 
$SU(2)_I$, vector-like doublet of PM fermion fields carrying $Y_I=0$, which is also an $SU(2)_L$ singlet; if these fields are also chosen to be color singlets then they will be assumed to 
have $Y/2=Q_{em}=-1$, analogous to the RH-leptons in the SM, \ie, 
\begin{equation}
{\cal E}_{L,R}=\begin{pmatrix} E_1 \\ E_2\\ \end{pmatrix}_{L,R}\,,
\end{equation}
which are clearly degenerate via $SU(2)_I$ and having a Dirac mass, $M_E$, that is assumed to be of order the $G_D$ breaking scale,  
\begin{equation}
M_E (\bar E_{1L}E_{1R} + \bar E_{2L}E_{2R})+{\rm h.c.}\,,  
\end{equation}
with $SU(2)_I$ here acting in the vertical direction. This mass term can, however, be augmented by also allowing a Yukawa coupling to the previously introduced $SU(2)_I$-breaking Higgs 
triplet, $\Sigma$, \ie,  
\begin{equation}
{\cal L}_\Sigma=y_\Sigma {\bar {\cal E}}_L \sigma \cdot \Sigma {\cal E}_R +{\rm h.c.}\,,
\end{equation}
with $\sigma$ being the set of Pauli matrices and $y_\Sigma$ an O(1) Yukawa coupling, so that when the vev become non-zero, the $E_{1,2}$ states now become split in mass, \ie,
\begin{equation}
M_{E_1,E_2}=M_E \pm  \frac{y_\Sigma v_T}{\sqrt 2}\,,
\end{equation}
but where we still expect that the ratio of these two masses to be of order unity. Note that this $2_I$ PM fermion doublet does {\it not} contain any SM fields so that, at least at this point, the 
$W_I^\pm$ gauge fields only connect $E_1-E_2$ and not, \eg, one of the $E_i$ with the SM $e_R$ as might have frequently occurred in our previously explored setups.  We note that, as the 
induced mass splitting between these two states, $\Delta=\sqrt 2 y_\Sigma v_T$, may possibly be larger 
than the mass of $W_I^\pm$ itself, $g_Iv_T$, we anticipate that the 2-body decay process $E_1\to E_2 W_I$ is also somewhat likely to be allowed to occur on-shell. 
Now, since the $E_i$ carry $T_{3L}=Y_I/2=0$ but still have $T_{3I}=\pm 1/2$ as well as $Y/2=-1$, due to their mass difference they can generate a finite and calculable, non-abelian KM at the 1-loop 
level between the SM hypercharge gauge boson, $\hat B$ and the $SU(2)_I$ gauge boson, $\hat W_{3I}$, with $T_{3I}=0$. Thus the set of kinetic terms in the Lagrangian for this model now 
contains an additional piece (suppressing Lorentz indices) of the form 
\begin{equation}
{\cal L}_{KM}=\frac{1}{2} {\bf {\hat \beta}} \cdot {\bf {\hat W_I}}{\hat B}\,,
\end{equation}
where the non-zero component of ${\bf {\hat \beta}}$ points in the `3' direction, projecting out $W_{3I}$; we see that no analogous abelian KM term is generated in this setup since the PM fields have 
$Y_I=0$. As is usual, since the magnitude of ${\bf {\hat \beta}}$ (which we'll just call $\beta$ from now on) is expected to be quite small, as we'll soon see below, this KM can be removed by the simple 
approximate field redefinitions 
\begin{equation}
{\hat B}\to B+\beta W_{3I}, ~~~~~~ {\hat W_{3I}} \to W_{3I}\,,
\end{equation}
while leaving the other fields unchanged. Hence we observe that the dark Higgs triplet field $\Sigma$ plays a very special role in this setup in that it not only is the source of the (partial) breaking of 
the $SU(2)_I$ gauge symmetries but also simultaneously leads to the generation of the non-abelian KM in this model by splitting the PM masses, which is itself $SU(2)_I$ breaking. Switching to 
the conventional normalization for the KM mixing parameter, $\beta \to \epsilon/c_w$, where $c_w=\cos \theta_w$, \etc, we in fact find that (assuming that no other PM fields are present) explicitly 
\begin{equation}
\epsilon=\Big(\frac{g_I}{e}\Big)~\Big(\frac{\alpha}{6\pi}\Big)~ln \Big(\frac{M_{E_1}^2}{M_{E_2}^2}\Big)\,,
\end{equation}
which is in the desired range, $\sim 10^{-4}-10^{-3}$ assuming $g_I/e$ is O(1). This is the simplest fermionic PM sector that one can imagine with PM fields in a non-trivial representation of 
$2_I$ which yields a finite and calculable value for the parameter $\epsilon$. 
Now it is possible that instead of the color and $SU(2)_L$ singlet VLF PM fields, $E_i$, we could have instead chosen (or added) color triplets which are also $SU(2)_L$ singlet VLF PM fields, 
$U_i$ and/or $D_i$. In such a case, essentially all of the analysis above would go through in exactly the same manner up to possible signs and factors of two arising in the expression for $\epsilon$. 
Thus the pure lepton-like case outlined here is seen to be just the simplest member of a class of such models.
Recalling that in the SM, $g_Y=g\tan \theta_w$, then in addition to conventional coupling to $g_IT_{3I}$, $W_{3I}$ will now pick up a KM-induced coupling to the SM hypercharge which, for 
convenience in later discussion, we may write suggestively in the form  
\begin{equation}
gt_w \frac{\epsilon}{c_w} \frac{Y}{2} W_{3I}\,,
\end{equation}
similar to what conventionally happens in the more familiar DP model setups. 

Before continuing there are several observations to make about this model. First, as we've already noted, the PM in this setup does not reside in a $2_I$ multiplet with any of the SM fields and is also 
vector-like with respect to $G_D$, as it already is with respect to the SM as usual; these two aspects are clearly linked. When any SM and PM fields both reside in the same $G_D$ 
representation they must share the same SM quantum numbers since by assumption $G_D$ commutes with the SM gauge group. However, this implies that if this multiplet is, \eg, 
left-(or right-)handed, then the same multiplet with the opposite handedness cannot simultaneously exist. If it did, it would require that the SM field contained within it to also be vector-like with 
respect to the $2_L1_Y$ gauge symmetries, something we know not to be the case for any of the usual fermions. Thus 
the only way for PM to be vector-like with respect to $G_D$ is for there not to be any SM fields sharing a common multiplet with it, as is the case here. Secondly, there is another possibility where 
an equally simple but distinct PM setup can be constructed: consider two vector-like, $2_I$ and $2_L$ PM singlets with different masses and having opposite values of $Y_I$ but with common values 
of the SM hypercharge. In such a case, the KM will be abelian, with the mixing now being between the two hypercharge gauge bosons, but the corresponding parameter, $\epsilon$, would remain 
finite and calculable. In some ways, such a setup would completely mirror the discussion below but with several important differences; however, such considerations are beyond the scope of the 
present work.

To go further with the present analysis,  we need to discuss the gauge symmetry breaking in some more detail. Subsequent to this high scale breaking, the SM itself will break in the familiar manner 
via the $\simeq 246$ GeV vev, $v_{SM}$, of an $2_L$ doublet (here with $Y/2=-1/2$) to which $W_{3I}$ now also couples via the above KM. Recalling that the SM $Z$ mass is just 
$M_{Z_0}=gv_{SM}/2c_w$ and that the SM photon field remains massless and decouples, the $2\times 2$ mass-squared matrix describing the mixing of the weak basis states $Z_0,W_{3I_0}$ can 
be written as 
\begin{equation}
{\cal M}^2_{ewk}=M_{Z_0}^2\begin{pmatrix} 1 & -\epsilon t_w\\ -\epsilon t_w& \epsilon^2 t_w^2\\  \end{pmatrix}\,,
\end{equation}
which is diagonalized at leading order in $\epsilon$ by the transformations
\begin{equation}
Z_0 \to Z+\epsilon t_w W_{3I}, ~~~~ W_{3I_0}\to W_{3I}-\epsilon t_w Z \,.
\end{equation}
This results in a tiny fractional shift in the SM $Z$ mass of $O(t_w^2\epsilon^2) \sim 10^{-8}$, which can be safely neglected, while $W_{3I}$ remains massless but picks up an additional 
mass-mixing induced 
coupling from the $Z$ so that the interaction of $W_{3I}$ with the SM fields in the mass eigenstate basis is now given by (here employing the SM relations $e=gs_w$ and $Q_{em}=T_{3L}+Y/2$) 
\begin{equation}
gt_w \frac{\epsilon}{c_w} \frac{Y}{2} +\epsilon t_w \frac{g}{c_w}\Big(T_{3L}-s_w^2Q_{em}\Big) =e\epsilon Q_{em}\,,
\end{equation}
in the usual DP-like fashion. Note that at this stage, while {\it both} the $W_{3I}$ and $B_I$ remain massless, it is only $W_{3I}$ that couples to the SM fields via KM. Any small masses for the light 
dark gauge bosons to be discussed below will not alter the coupling above up to very small mass suppressed correction terms. However, further symmetry 
breaking at lower scales will generally lead to $W_{3I}-B_I$ mixing so that both of the resulting mass eigenstates will experience KM with the SM photon as we will now discuss. 

At this stage, the remaining unbroken dark gauge symmetries are just $U(1)_{T_{3I}}\times U(1)_{Y_I}$ and to completely break these we need to employ (at least) two dark Higgs fields, $H_{1,2}$, 
which will obtain the vevs, $v_{1,2}$, and will also supply us with the Goldstone bosons needed for both the $W_{3I}$ and $B_I$ (or some linear combinations thereof) to obtain masses. Subject to 
the constraint that both gauge symmetries are indeed broken, there still remains some significant freedom in the choice of quantum numbers for these two Higgs fields. Here we will 
be guided by other plausible requirements based on satisfying phenomenological and experimental constraints. Note that these small vevs appearing in $H_{1,2}$ will also naturally lead to the 
existence of two light dark Higgs, $h_{1,2}$. In practice these are approximately linear combinations of 
the light CP-even states originating from the representations $H_{1,2}$ whose masses, in the appropriate limit, are roughly set by these same vevs. 

The first of these requirements, $(i)$, arises from a simple observation. As noted above, unlike in other previously examined PM models, the $W_I^\pm$ here do not connect the PM fields with SM 
ones which have identical SM quantum numbers. Thus, while $E_1$ can decay to $W_IE_2$, the lighter $E_2$ is at this point a charged, stable field with a $\sim$ TeV scale mass, which is 
not phenomenologically acceptable. This problem can 
be alleviated by choosing, \eg, $H_1$ to be an $2_I$ doublet with $Y_I=0$, which then obtains the vev, $v_1 \lsim 1$ GeV. One can then couple ${\cal E}$ to the SM right-handed charged 
lepton with the same SM quantum numbers, generically described as $e_R$, via the $H_1$, \ie, 
\begin{equation}
{\cal L}_{mix}=y_{mix} {\bar {\cal E}}_L e_RH_1 +{\rm h.c.}\,.
\end{equation}
Then, \eg, the fields $E_2-e$ will now experience a small mass mixing once $H_1$ obtains its vev thus allowing for the decay $E_2\to e_RW_{3I}$ (which is always kinematically allowed) as a 
FCNC-like 
process. Given the relative mass scales involved, we might expect the relevant mixing angle to be $\phi_{eE}\sim v_1/v_T \sim O(\epsilon)$ which is sufficient for this decay to the avoid the familiar 
experimental lifetime constraints on new heavy charged states.  This is especially so due to the longitudinal couplings of the the mass eigenstates formed from both $W_{3I}$ and $B_I$ after 
symmetry breaking, leading to a significant enhancement in this amplitude which can compensate for this small mixing as has been discussed in earlier work\cite{Rizzo:2018vlb,Rizzo:2022qan}.  
This is even more obvious if we take $y_{mix}$ to be an O(1) coupling and consider the Goldstone component in $H_1$ and its coupling to, \eg, $\bar E_L e_R$. 
We also not that the heavier 
$E_1$ PM state will also obtain an additional decay channel, $E_1\to e_R W_I$, as result of this same mixing process; this may or may not compete with the already discussed $E_1 \to E_2W_I$ 
decay depending upon the exact mass eigenvalues and the size of the mixing. Note that if we had instead gone the route of color triplet VLFs as PM, this same mixing mechanism 
would be required to allow the lighter of the two quark-like states to mix with the right-handed SM $u$ or $d$ quarks; the quantum numbers of the $H_1$ in such a setup would be exactly the same as 
chosen in the present case.

Next, let us consider the DM sector of our setup which we will assume here to be fermionic. Then, as is well-known and as was briefly discussed in the Introduction, the easiest way in which to avoid 
numerous constraints arising from, \eg, the CMB as well as direct and indirect detection DM searches, is to require that, ($ii$), the DM, $\chi$, be pseudo-Dirac due to the generation of Majorana 
mass terms as well as a bare Dirac mass producing an inelastic DM-type scenario. These will split $\chi$ into two states, $\chi_{1,2}$, now having masses $m_1<m_2=m_1(1+\delta)$, which 
clearly avoids direct detection for sufficiently large values of $\delta$ (which we will assume is the case in the analysis below), since the coupling to the responsible gauge boson(s) 
is off-diagonal at tree-level. In such a case, the DM will achieve its relic abundance via co-annihilation into SM final states 
which becomes very highly suppressed at later times as the temperature falls significantly, thus avoiding CMB constraints as well as present day indirect detection searches. 
To do this in the simplest possible manner, we imagine that $\chi$ lives in some vector-like representation, $X_{L,R}$, of $G_D$ in such a way that when $H_2$ develops a vev, these Majorana 
mass terms are generated; in this case we can write a general set of mass terms in the Lagrangian as
\begin{equation}
{\cal L}_{Dark} = m_X\bar X_L X_R +\frac{1}{2}(y_L \bar X_L^cX_L+y_R\bar X_R^cX_R)H_2+{\rm {h.c.}}\,,  
\end{equation}
and here, for simplicity, we will assume the equality of the left and right Yukawa couplings, $y_L=y_R=y$.  Now if the DM were to be in an $SU(2)_I$ singlet representation, $H_2$ would also need to 
be a singlet with $Y_I \neq 0$. In such a case, $W_{3I}$ and $B_I$, would each obtain masses from one of the vevs of $H_{1,2}$, but we would then be faced with the unhappy situation that these 
states would never mix and so already be mass eigenstates with $B_I$ still uncoupled to the fields of the SM by KM.  The next simplest possibility, one we considered earlier\cite{Rizzo:2024bhn} and 
which we will adopt here, is that $X$ is instead an $2_I$ doublet with $Y_I/2=1/2$ and where $\chi$ itself has $T_{3I}=1/2$, \ie, 
\begin{equation}
X_{L,R}=\begin{pmatrix}\chi \\ \psi \\ \end{pmatrix}_{L,R} \,.
\end{equation}
In such a case, $H_2$ must be identified as an $SU(2)_I$ triplet with $Y_I/2=1$ to generate the appropriate additional mass terms once it obtains a vev. While the associated phenomenology 
of the DM sector in this case will be discussed in the next Section, these two Higgs fields and their corresponding vevs will be sufficient for our purposes. However, we note that it is also possible 
to go further and also couple the doublet $X_{L,R}$ to the triplet $\Sigma$ already introduced above in such a way as to split the originally degenerate Dirac masses, $m_X$, of the $\chi, \psi$ 
doublet members, \ie, 
\begin{equation}
{\cal L}_\Delta'=y'_\Sigma {\bar X}_L \sigma \cdot \Sigma X_R +{\rm h.c.}\,.
\end{equation}
If we assume that this Yukawa coupling, $y'_\Sigma$, has a magnitude somewhat similar to those of the first or second generation fermions to the SM Higgs, then $\psi$'s Dirac mass (which is the 
only mass that it possesses) can be significantly increased relative to that of $\chi$, allowing for a decay with a relatively short decay lifetime.  From now on we will denote the smaller Dirac mass 
of the DM state, $\chi$, by $m_D$ for clarity. 

We note that while these requirements above are strongly suggestive, they are not the unique choice for the set of Higgs fields that break the low energy gauge symmetries and other possibilities 
may be contemplated. We will, however, adopt these specific representations and associated quantum numbers for their vevs in the discussion that follows.  

Putting the requirements ($i$) and ($ii$) together from our discussion above, the resulting $W_{3I}-B_I$ mass-squared matrix that arises from these two vevs can be written as 
\begin{equation}
{\cal M}^2=\begin{pmatrix} M_A^2+M_B^2 & t_IM_B^2\\ t_IM_B^2& t_I^2M_B^2\\  \end{pmatrix}\,,
\end{equation}
where we have defined the ratio of gauge coupling $g_{Y_I}/g_I=t_I>0$ in analogy with the SM, and for simplicity, we have further defined the mass-squared parameters
\begin{equation}
M_A^2=(g_Iv_1)^2/2,~~~~M_B^2=(g_Iv_2)^2\,.  
\end{equation}
Introducing the dimensionless ratio $\rho=M_A^2/M_B^2$, this mass-squared matrix can be diagonalized via a rotation through an angle, $\phi$, given by 
\begin{equation}
\tan 2\phi =\frac{2t_I}{\rho+(1-t_I^2)}\,,
\end{equation}
yielding the mass-squared eigenvalues, $M_{1,2}^2=M_B^2\lambda_\mp$ (\ie, $M_1<M_2$), where $\lambda_\mp$ are given by the expression
\begin{equation}
\lambda_\mp=\frac{1}{2}\Big[(1+t_I^2)+\rho\Big]\mp \frac{1}{2}\Big[(1+t_I^2)^2+\rho^2+2\rho (1-t_I^2)\Big]^{1/2}\,.
\end{equation}
Interestingly, we see that when $M_A^2 \to 0$ (\ie, $\rho \to 0$) then $\phi \to \theta_I$ and the eigenstate with a non-zero mass in this limit appears as the conventional DP which couples to the usual 
combination $Q_D=T_{3I}+Y+I/2$; of course the remaining gauge field is seen to be massless in this same limit. We will refer to these gauge boson mass eigenstates more generally as 
$Z_{1,2}$ in the discussion that follows.
Continuing, we see that, since $W_{3I}$ is just a linear combination of these two mass eigenstates, \ie, $W_{3I}=Z_1 c_\phi+Z_2 s_\phi$, both $Z_{1,2}$ will now have KM-induced couplings to SM 
fields which takes the familiar form
\begin{equation}
e\epsilon Q_{em}(c_\phi Z_1+s_\phi Z_2)= v_1^{SM}Z_1+v_2^{SM}Z_2\,,
\end{equation}
while the general gauge couplings of the $Z_{1,2}$ to dark sector fields and PM can now be written as 
\begin{equation}
g_I[c_\phi T_{3I}-s_\phi t_I (Y_I/2)]Z_1+g_I[s_\phi T_{3I}+c_\phi t_I (Y_I/2)]Z_2\,,
\end{equation}
and we can easily check that when $\rho \to 0$ the $Z_2$ couplings becomes just those of the familiar DP.

Now for later convenience, let us define the ratio of these physical gauge boson masses by the parameter $\lambda_R^2=\lambda_+/\lambda_-=M_2^2/M_1^2$ which is $>1$ by definition. Using the 
two invariants of the ${\cal M}^2$ matrix, it is then straightforward to obtain a bound on $t_I$ as a function of $\lambda_R$ by demanding that the ratio $\rho$ be a real quantity assuming that both 
$t_I$ and $\lambda_R$ are used as physical input parameters (as they are linked to observables). To see this, we first define the dimensionless ratio of the matrix invariants 
\begin{equation}
\Lambda=\frac{Det ~{\cal M}^2}{(Tr ~{\cal M}^2)^2}=\frac{\rho t_I^2}{[\rho +(1+t_I^2)]^2}=\frac{\lambda_R^2}{(1+\lambda_R^2)^2} <\frac{1}{4}\,,
\end{equation}
and then solve for $\rho(\Lambda)$ requiring a physical, \ie, a real-valued result. From these considerations, we can then obtain a lower bound on the value of $t_I$ as a function of $\lambda_R$, 
\ie,
\begin{equation}
t_I^2> \frac{4\Lambda}{1-4\Lambda} ~\to ~t_I > \frac{2\lambda_R}{\lambda_R^2-1}\,,
\end{equation}
which we will make use of in the analyses that follow. It is important to note that this lower bound weakens reasonably rapidly as $\lambda_R^2$ increases.

\section{Phenomenological Implications: DM at Low Energies with Two $U(1)$'s}

As discussed above, the combination of the Dirac mass and the dark Higgs triplet-induced Majorana mass term for $\chi$ splits this field into two pseudo-Dirac states, $\chi_{1,2}$, having masses 
$m_{1,2}=m_D\mp yv_2/\sqrt 2$. As is well-known and as used in our previous work\cite{Rizzo:2024bhn}, these states will interact with the $Z_{1,2}$ gauge bosons in an off-diagonal manner, \ie, via 
\begin{equation}
{\cal L}_{int}=\frac{i g_I}{2} [\bar \chi_2\gamma_\mu \chi_1 -(1\leftrightarrow 2)] (v^D_1Z_1+v^D_2Z_2)\,,
\end{equation}
where the parameters $v^D_i$ are directly obtainable from the general coupling expressions as presented by Eqs.(22) and (23) above:
\begin{equation}
v^D_1=\frac{1}{2}(c_\phi-t_Is_\phi),~~~~~v^D_2=\frac{1}{2}(s_\phi + t_I c_\phi)\,.
\end{equation}
Given some input values for the ratio of the lightest (stable) DM state to that of $Z_1$, $r=m_1/M_1$, the relative mass splitting between $\chi_2$ and $\chi_1$, $m_2/m_1=1+\delta $, the ratio of 
the squares of the dark gauge boson masses, $\lambda_R^2$, as well as the value of $t_I$, we can determine the DM co-annihilation cross section for, \eg, $\chi_1\chi_2 \to e^+e^-$ in a rather 
straightforward 
manner up to an overall unknown coupling factor. Note that in the symmetric $y_L=y_R=y$ limit considered here that there are no `diagonal' annihilation channels into SM final states, \eg, 
$\chi_1\chi_1, \chi_2\chi_2 \to e^+e^-$.  Further, it will be assumed that any light dark Higgs fields that are present in this setup do not contribute significantly to this cross section as their couplings to 
the light SM fermions can only arise via mixing with the SM Higgs, where they are already found to be small due to the appearance of ratios $m_{e,\mu}/v_{SM}$ in these couplings. This mixing of 
the light dark Higgs with the SM Higgs is also found to be quite highly constrained by the observed limit on the $H(125)$ branching fraction into invisible final states, $B_{inv}< 0.10-0.15$, by 
both ATLAS and CMS at the LHC\cite{ATLAS:2023tkt,CMS:2023sdw}. Correspondingly, in the case where the light gauge bosons may decay to leptons there are also other constraints as well 
from searches involving collimated lepton pairs\cite{ATLAS:2024zxk,CMS:2024jyb} which will also enforce a strong suppression upon this mixing. 
We will, however, have a few more brief words to say about the possible role of these light dark Higgs fields in our later discussions below. 

Before continuing, we recall that the value of $r=m_1/M_1$, is bounded from above in these setups. When $r>1$, the $s$-wave, on-shell process, $\bar \chi_1 \chi_1 \to 2 Z_1$, $Z_1 \to$ SM, 
becomes kinematically allowed and thus significantly violates the CMB bound on the deposition of electromagnetic energy into the thermal bath when $Z_1$ decays to, say, $e^+e^-$. To safely 
avoid this mishap we need not only require that $r<1$ but an even stronger condition is required since the collision of the two $\chi$'s, with both being towards the top of their thermal energy 
distributions, can lead to a center of mass energy, $\sqrt s$, somewhat larger than just the naive value of $2m_1$. While the actual upper bound depends in a complex manner on of the interplay 
of the model parameters, here we will for simplicity only require that $r<0.9$\cite{Rizzo:2020jsm}. In a similar vein, 
we'll also require that small values of $r$ are excluded, roughly corresponding to excluding very light DM masses less than 
$\simeq 10$ MeV to avoid any constrains coming from the BBN\cite{Giovanetti:2021izc,Chu:2022xuh,Sabti:2021reh,Sabti:2019mhn}; in practice, we will simply require $r>0.1$.  Lastly, we'll also 
assume that $\lambda_R^2$ is sufficiently great so that the relativistic Breit-Wigner approximation remains valid{\footnote {See, for example, \cite{Rizzo:2012rb} and references therein.}} in that 
the mass separation between the $Z_i$ resonances is always large compared to their individual total widths.

\begin{table}
\caption{Benchmark Points}\label{qtab}
The benchmark set of pairs of $\lambda_R^2$ and $t_I$ parameter values employed in the relic density calculation described in the text with the corresponding results shown below in 
Fig.~\ref{fig1}-\ref{fig5}. The value of the ratio $R_2$ showing the effect of the two gauge boson exchanges in the 3-body decay of $\chi_2 \to \chi_1 e^+e^-$ given in Eq.(43) assuming $f=1$ is 
also given.
\begin{center}
\begin{tabular}{ l c c }
\hline
\vspace{-0.1cm}\\
$\lambda_R^2$ &  $t_I$   & $R_2$\vspace{0.1cm}\\
\hline 
\vspace{-0.05cm}\\
$3 $& 1.8 & 0.55 \vspace{.01cm}\\
$4$ & 1.5  & 1.14\vspace{.01cm}\\
$5$ & 1.2 & 4.54 \vspace{.01cm}\\
$6$ & 1.0  & 1.56\vspace{.01cm}\\
$7$ & 0.90 &1.33 \vspace{.01cm}\\
$8$ & 0.85  &1.30\vspace{.01cm}\\
$9$ & 0.80  & 1.23\vspace{.01cm}\\
\hline 
\vspace{0.05cm}\\
\end{tabular}
\end{center}
\end{table} 

To begin, we recall that for co-annihilating DM, where the $\chi_1 \chi_1$ and $\chi_2\chi_2$ annihilation cross sections, $\sigma_{11,22}$, can be neglected, we may write the effective cross section 
weighted by the relative velocity as 
\begin{equation}
\sigma v_{rel}= {\cal F}^2~\frac{2R{\sigma_{12}}}{1+R^2}\,,
\end{equation}
where $\sigma_{12}$ is the `parton-level' velocity-weighted $\chi_1 \chi_2 \to$ SM cross section which occurs via the $Z_{1,2}$ exchanges, to be described below,  $R$ is given by the expression 
\begin{equation}
R=(1+\delta)^{3/2}e^{-\delta x_F}\,,
\end{equation}
with $m_2/m_1=1+\delta$, as defined above and, as usual, $x_F=m_1/T_F \simeq 20$, with $T_F$ being the thermal freeze out temperature. Finally, the overall rate for this process is set by the 
model-dependent numerical pre-factor, ${\cal F}$, given by  
\begin{equation}
{\cal F} =\Big(\frac {\epsilon g_{eff}}{10^{-4}}\Big)~ \Big(\frac{100 ~{\rm MeV}}{M_{Z_1}}\Big)\,,
\end{equation}
where we have defined the effective coupling combination 
\begin{equation}
g_{eff}= g_Ic_\phi v^D_1=\frac{1}{2}g_I c_\phi (c_\phi-t_I s_\phi)\,.
\end{equation}
Once the thermally averaged value for $\sigma v_{rel}$, \ie, $<\sigma v_{rel}>$,  is determined by integration over the DM thermal distributions and then compared with that required to obtain the 
observed\cite{Planck:2018vyg} DM relic abundance in this case, $\sigma_0 \simeq 4.4\times 10^{-26} \rm {cm^3 sec^{-1}}$\cite{Steigman:2012nb,Steigman:2015hda,Saikawa:2020swg}, 
we can then extract the corresponding value needed for the quantity ${\cal F}$ as a function of the other model parameters. The particular parameter space 
regions where ${\cal F}$ is not required to be too large (and hence more likely to already be excluded by other experiments) would then clearly be seen as the more favorable and we will employ this 
perspective below.

To proceed further, we need to have an expression for $\sigma_{12}$ before it can be integrated over the thermal distributions of the $\chi$'s. In the CoM frame we find that this is given by 
\begin{equation}
\sigma_{12}=\frac{2\alpha_{em}}{s} ~(\epsilon g_{eff})^2~\Bigg[\frac{m_1m_2+e_1e_2+\frac{p^2}{3}}{s}\Bigg]~\sum_{ij} ~\Big(P_{ij}~\tilde v^D_i \tilde v^D_j \tilde v^{SM}_i \tilde v^{SM}_j \Big)\,,
\end{equation}
with $e_{1,2}$ (here in lower case to avoid any  confusion with the fermion PM fields introduced above) and $p$ being the energy and momenta of the colliding $\chi_{1,2}$ particles which we can 
express in terms of $s$ and the $\chi_{1,2}$ masses as 
\begin{equation}
e_1=\frac{s+m_1^2-m_2^2}{2\sqrt s},~~e_2=e_1 (1\leftrightarrow 2), ~~ p_1=p_2=p=(e_i^2-m_i^2)^{1/2}\,,
\end{equation}
and where, since we have pulled outside of this expression into the definition of ${\cal F}$ both $g_{eff}$ and $\epsilon$, one finds that $\tilde v^D_1 = \tilde v^{SM}_1 =1$, while 
\begin{equation}
\tilde v^D_2= \frac{t_\phi+t_I}{1-t_\phi t_I} =\tan(\phi+\theta_I),~~~ \tilde v^{SM}_2= t_\phi\,.
\end{equation}
Also making an appearance in this cross section expression is the familiar propagator factor for the $Z_i$ exchanges: 
\begin{equation}
P_{ij}= s^2~ \frac{(s-M_i^2)(s-M_j^2)+(M_i\Gamma_i)( M_j\Gamma_j)}{[(s-M_i^2)^2+(\Gamma_i M_i)^2)][(1\to 2)]}\,,
\end{equation}
where the $\Gamma_i$ are the total widths of the $Z_i$ which are calculable after making a few assumptions. For example, when $m_1+m_2<M_i$, the $Z_i$ will almost exclusively decay into 
$\chi_1\chi_2$ since decays to the SM final states, such as $e^+e^-$, are always suppressed by an additional factor of $\epsilon^2$. In such a case, to an excellent approximation one has 
\begin{equation}
\frac{\Gamma_i}{M_i} = \frac{g_I^2(v^D_i)^2}{24\pi} \Big[(1+x_a-x_b)^2-4x_a\Big]^{1/2} \Big[2-x_a-x_b-(x_a-x_b)^2+6(x_ax_b)^{1/2} \Big]\,,
\end{equation}
with $x_{a,b}=m^2_{a,b}/M^2_i$.
Whereas, if the $\chi_1 \chi_2$ final state is {\it not} kinematically accessible and $Z_i$ decays only to, \eg, the SM (massless) $e^+e^-$ final state were to be allowed (which we will assume to be 
the case here), we would instead obtain
\begin{equation}
\frac{\Gamma_i}{M_i} = \epsilon^2 ~\frac{\alpha_{em} (v_i^{SM})^2}{3}\,.
\end{equation}
These widths are clearly directly connected with the $r<0.9$ constraint discussed above together with the chosen values of $\delta$ as well as $\lambda_R^2$ in the case of $Z_2$, as will be 
seen by the numerical results to be presented below. It should be noted that in almost all of the cases to be discussed, this mass constraint will allow the $Z_2$ to almost always decay to 
$\chi_1 \chi_2$ on-shell as the kinematic requirement for this, $r(2+\delta)<\lambda_R$, is mostly always met. Assuming no other decay modes for $Z_2$ are allowed, this lone mode implies that 
the $Z_2$'s decay will almost always be invisible unless the subsequent decay of $\chi_2\to \chi_1 e^+e^-$, which, as will be discussed below, now occurs via virtual $Z_i$ exchanges, is visible 
However, as is well known this rate is suppressed by both $\epsilon^2$ as well as by the 3-body phase space, so likely is not sufficiently rapid to be experimentally observed above backgrounds 
especially when it is highly boosted as it may be at colliders.

As noted, in the numerical analysis that follows it will be assumed that no other decay channel are open, especially for the $Z_2$; however, other possibilities do exist. 
For example, since the vevs of the two light dark Higgs representations induce 
mixing between the $W_{3I}$ and $B_I$ into the $Z_i$ mass eigenstates, it is natural that there remain off-diagonal couplings between these gauge bosons and the two corresponding light 
Higgs scalars, $h_{a=1,2}$, of the generic form
\begin{equation}
{\cal L}_{o.d.}=g_I\mu_a M_2 Z_2^\nu Z_{1\nu}h_a\,
\end{equation}
where the $\mu_a$ are dimensionless parameters that are rather complex combinations of the model couplings, \etc,  but which are generally of order unity, resulting from the required mixings in both 
the gauge and scalar boson sectors to obtain the relevant mass eigenstates. Defining 
$r_1=1/\lambda_R^2=M_1^2/M_2^2$ and  $r_a=m_{h_a}^2/M_2^2$, one finds that the partial decay width for the process $Z_2\to Z_1h_a$ (assuming it to be kinematically allowed) to be given by 
\begin{equation}
\frac{\Gamma_2}{M_2} = \frac {g_I^2\mu_a^2}{192\pi}~ \frac{1}{r_1}~\big[\lambda(1,r_1,r_a)+12r_1\big]~\lambda^{1/2}(1,r_1,r_a)\,,
\end{equation}
with $\lambda$ here being the familiar kinematic function
\begin{equation}
\lambda(1,r_1,r_a)=(1+r_1-r_a)^2-4r_1\,.
\end{equation}
In principle, especially when $\lambda_R^2$ becomes somewhat large, we see that this partial width may compete with or even exceed that for $Z_2 \to \chi_1\chi_2$. However as noted, in the 
numerical study that follows we will ignore the possibility that this decay mode may be available for simplicity though we should be mindful that it may be of phenomenological relevance in some 
regimes of parameter space.

\begin{figure}[htbp]
\centerline{\includegraphics[width=5.0in,angle=0]{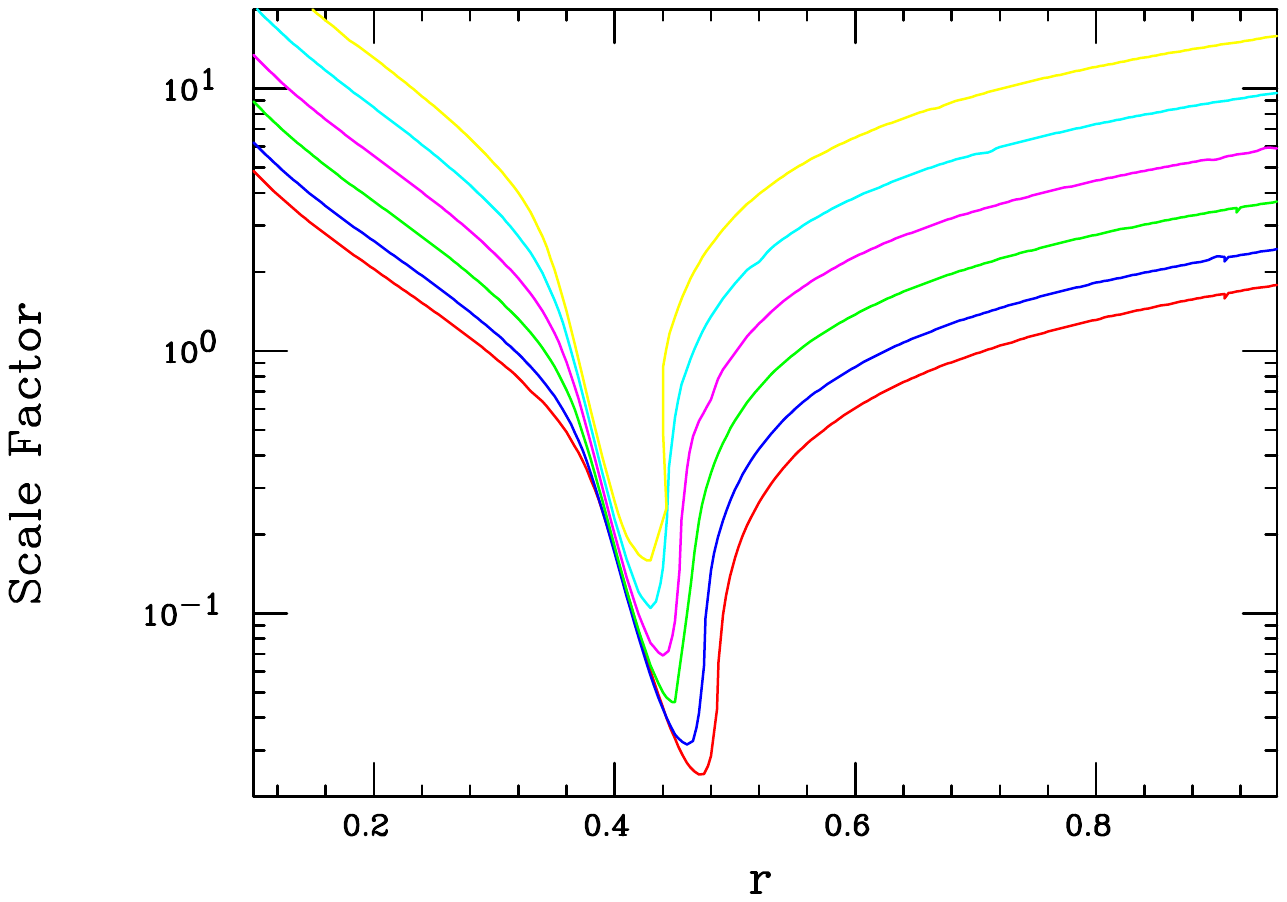}}
\vspace*{-0.8cm}
\centerline{\includegraphics[width=5.0in,angle=0]{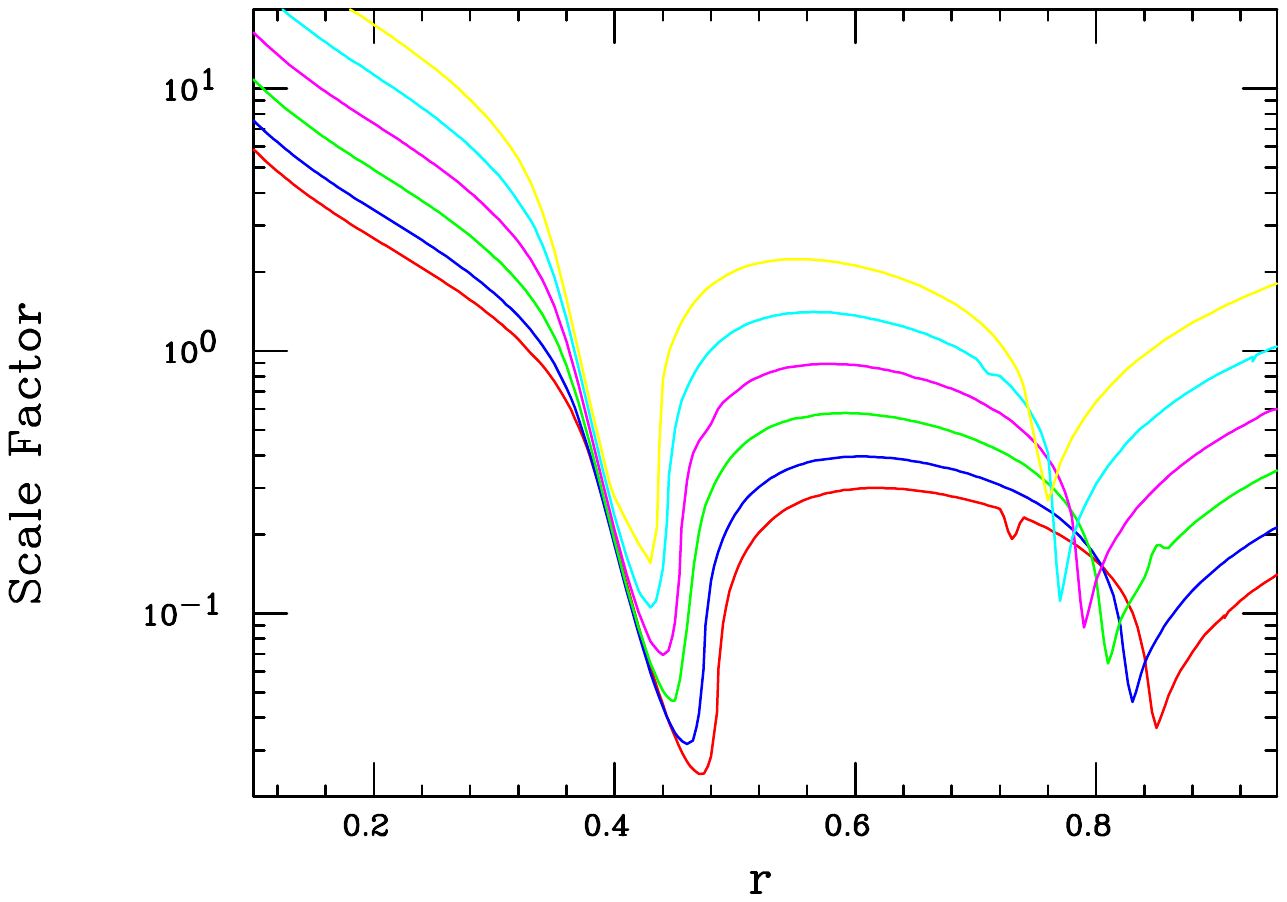}}
\vspace*{-1.3cm}
\caption{Value of the coupling scaling factor, ${\cal F}$, as defined in the text, as a function of $r$ for, from bottom to top on the right,  $\delta=0.05-0.30$ in steps of 0.05. The (Top) panel assumes 
the familiar case of a single light dark gauge boson whereas the (Bottom) panel assumes two gauge bosons with $\lambda_R^2=3$ and $t_I=1.8$ for purposes of demonstration.}
\label{fig1}
\end{figure}

Given $\sigma_{12}$, expressed as a function of the dimensionless parameters $r,\delta, t_I$ and $\lambda_R^2$, we can perform the thermal averaging, compare with the necessary cross section to 
achieve the observed DM relic abundance and extract the required value of ${\cal F}$. However, if we are to explore the effect of now having two light dark gauge bosons instead of just the usual DP, 
in particular, the effects associated with their interference (\eg, the so-called Bactrian Effect\cite{Rizzo:2021pxo}), we first must have 
as a reference the `standard' result for the case where the second gauge boson decouples in the limit when $M_2^2\to \infty$ with which to compare. Again, it will be assumed for simplicity that the 
only SM final state available is just $e^+e^-$. This is shown in the top panel of 
Fig.~\ref{fig1} as a function of $r$ assuming different values of the mass splitting parameter $\delta$. Clearly, this result is, by definition, independent of both $t_I$ and $\lambda_R^2$. As we can 
see from this Figure, (the square root of) the `inverse' of the usual Breit-Wigner peak shows itself via the minimum in the required value of ${\cal F}$ but which 
is distorted asymmetrically around this peak (seen as the  minimum) by the velocity distribution weighting of the cross section; clearly the region around the minimum is `preferred'. The width of 
this special region and the 
shallowness of the distribution at the minimum is significantly affected by the chosen value of $\delta$.  As $\delta$ increases, the sum of the $\chi_{1,2}$ masses also increases pushing the 
minimum to lower $r$ values. Also as $\delta$ increases, the required value of ${\cal F}$ needs to increase to compensate for the relative exponential suppression in the $\chi_2$ thermal distribution. 
If, \eg, we demanded values of ${\cal F}$ less than $\simeq$ unity, then we need either require that $r \sim 0.3-0.4$ or $r \sim 0.44-0.7$ with the tighter constraint applying when $\delta$ obtains 
larger values. Certainly, the region above and significantly far away from the minimum appears somewhat disfavored (especially so for higher $\delta$ values) given that larger values of ${\cal F}$ 
seem to be necessary in such cases.

In order to explore how these quite familiar results will change due to the existence of the $Z_2$ and, in particular, $Z_1-Z_2$ interference effects, we will consider the gauge boson mass ratio range 
$3\leq \lambda_R^2\leq 9$ beyond which we can easily imagine how the asymptotic region, essentially shown in the top panel of Fig.~\ref{fig1}, is eventually more closely approached. This analysis 
requires choosing a set of values for $t_I$ satisfying the lower bound given in Eq.(25); these benchmark values are given in Table~\ref{qtab} and are generally chosen to lie not too far above 
this lower bound. We expect that almost all of the modifications due to the added $Z_2$ exchange will take place above the $Z_1$ and this will be roughly borne out in the analysis that follows.  
As is easily  seen, at the amplitude level, the magnitude of the interference of the $Z_{1,2}$ contributions, be it constructive or destructive, is essentially controlled by both the separation between 
the two gauge boson masses via $\lambda_R^2$ as well as the both the sign and the magnitude of ratio of the product of their couplings to the initial and final states, \ie,
\begin{equation}
{\cal R}=\frac{v_2^{SM}v_2^D}{v_1^{SM}v_1^D}= t_\phi t_{\phi+ \theta_I}\,,
\end{equation}
which can be uniquely determined once $t_I$ is known for a fixed value of $\lambda_R^2$ as was discussed in the previous Section. 

\begin{figure}[htbp]
\centerline{\includegraphics[width=5.0in,angle=0]{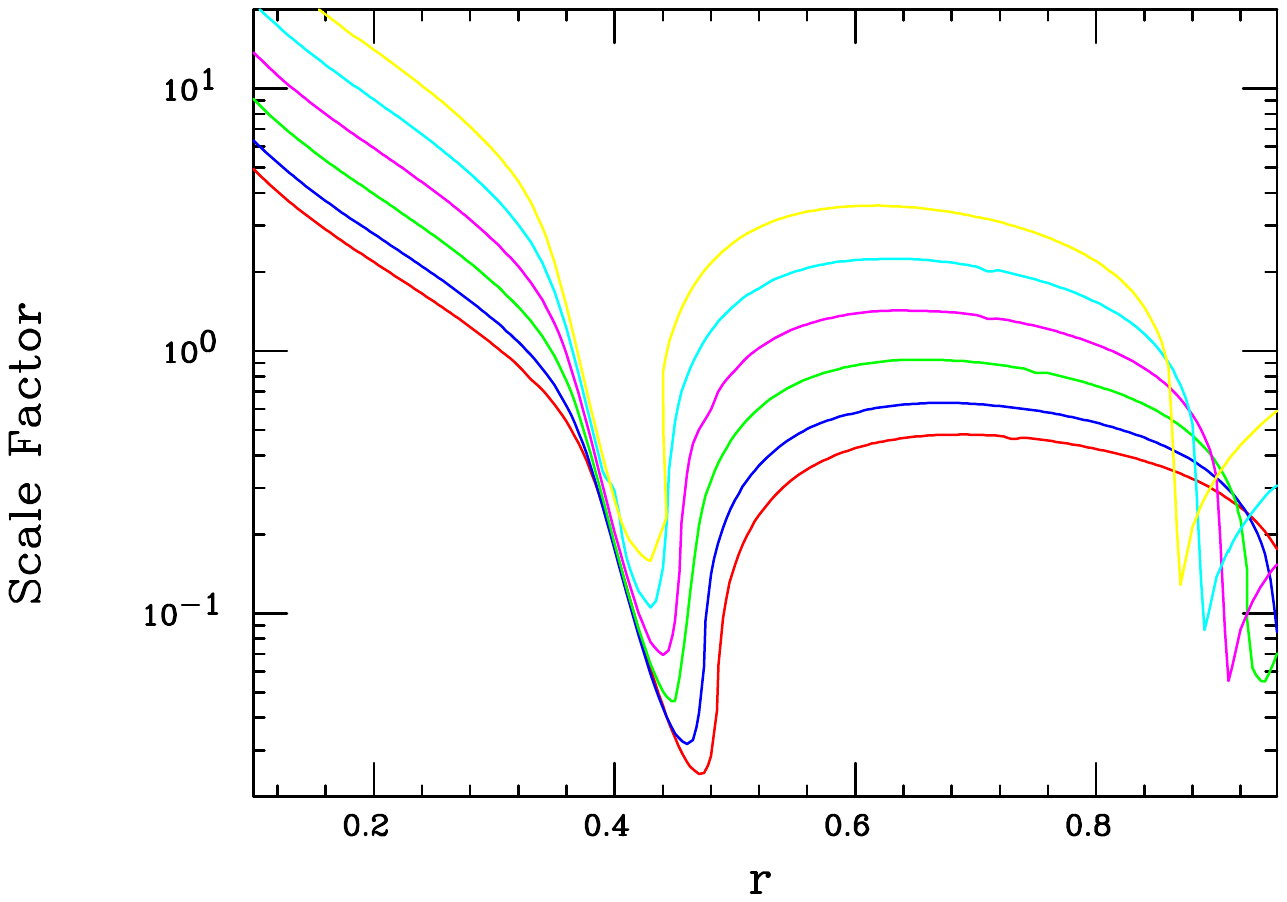}}
\vspace*{-0.8cm}
\centerline{\includegraphics[width=5.0in,angle=0]{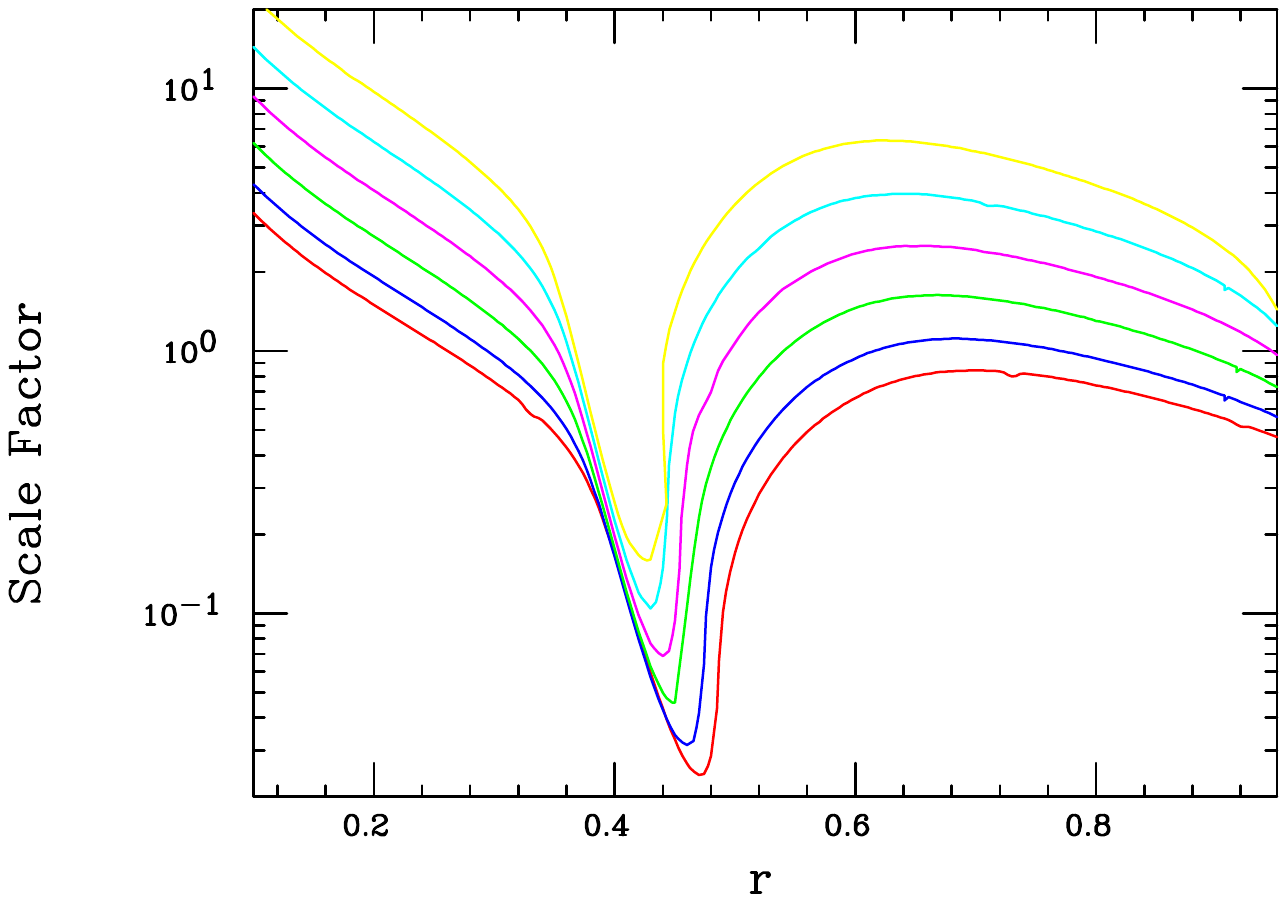}}
\vspace*{-1.3cm}
\caption{Same as the bottom panel in the previous Figure but now for (Top) $\lambda_R^2=4$ with $t_I=1.5$ and (Bottom) $\lambda_R^2=5$ with $t_I=1.2$.}
\label{fig2}
\end{figure}

Let us begin with the smaller values of $\lambda_R^2$ and then move upwards towards larger values. In the lower panel of Fig.~\ref{fig1} we consider the case of benchmark point 
$\lambda_R^2=3, t_I=1.8$ where several things are immediately apparent. First, and 
most obvious, is the presence of the second minimum near $r\sim 0.8$ due to the existence of the relatively close by $Z_2$. The fact that this minimum is visible in the range $r<0.9$ indicates that the 
kinematic requirement $r(2+\delta)<\lambda_R$ needed for the on-shell decay $Z_2\to \chi_1 \chi_2$ to be allowed on-shell is not always satisfied once $r$ is larger than $\simeq 0.75$.  Second, 
we see that in the region between the two resonances as well as just beyond the $Z_2$ minimum, the required value of ${\cal F}$ is always less than $\simeq 1-2$ due to the (constructive) interference 
of the two amplitudes making this entire parameter region now no longer `disfavored' and certainly, at the very least, more viable thus opening up more of the parameter space for further study. 
Third, the effect of the $Z_2$ exchange is not very noticeable in 
the region at or immediately below the $Z_1$ resonance minimum (in comparison to the case of just a single exchange) as might have been expected due to the proximity of this resonance 
although some modest influence of destructive interference is noticeable under somewhat closer scrutiny for sufficiently small values of $r\lsim 0.3$. The bottom line is that we see that the 
constructive interference due to having two nearby resonances opens up a significant region of parameter space that was previously unfavorable. Furthermore, in this region, the $Z_1$ is seen 
to always decay to SM final states.  It is to be noted that the minimum corresponding to the $Z_2$ is somewhat more shallow than that due to $Z_1$. Since the couplings are comparable, this is 
mainly due to the fact that $Z_1$ is generally far more narrow as only decays to SM final states (here, just $e^+e^-$) are likely allowed whereas the $Z_2$ can also decay to the $\chi$'s and so is much 
wider thus suppressing the height of the cross section peak. Finally, in the region {\it above} the second resonance and not easily seen in this Figure (but would be if, \eg, values of $\lambda_R^2=2$ 
were to be entertained), the two amplitudes will destructively interfere forcing us to chose even larger values of ${\cal F}$ than in the single exchange model to obtain the observed DM relic density. 
This region continues to be disfavored.

\begin{figure}[htbp]
\centerline{\includegraphics[width=5.0in,angle=0]{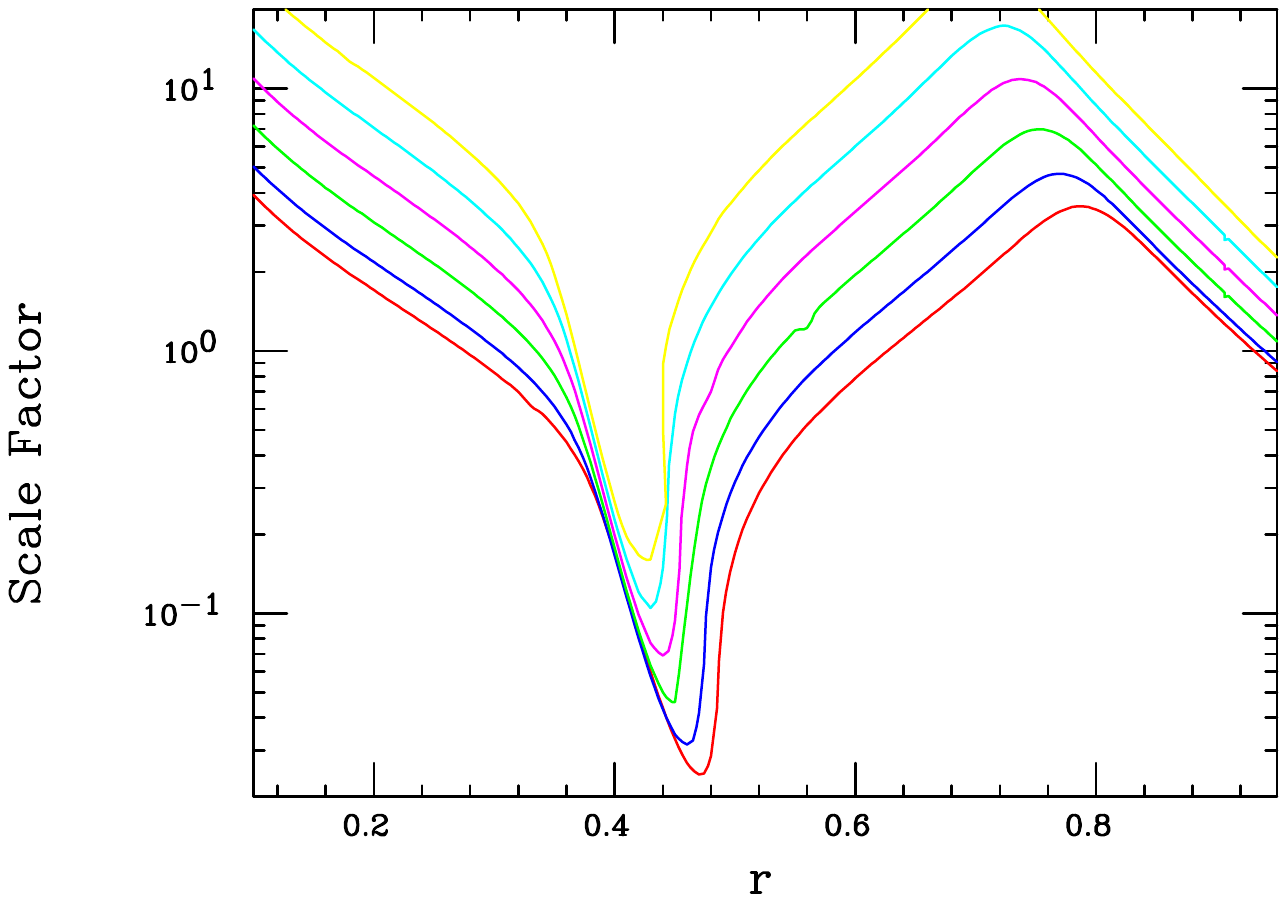}}
\vspace*{-0.8cm}
\centerline{\includegraphics[width=5.0in,angle=0]{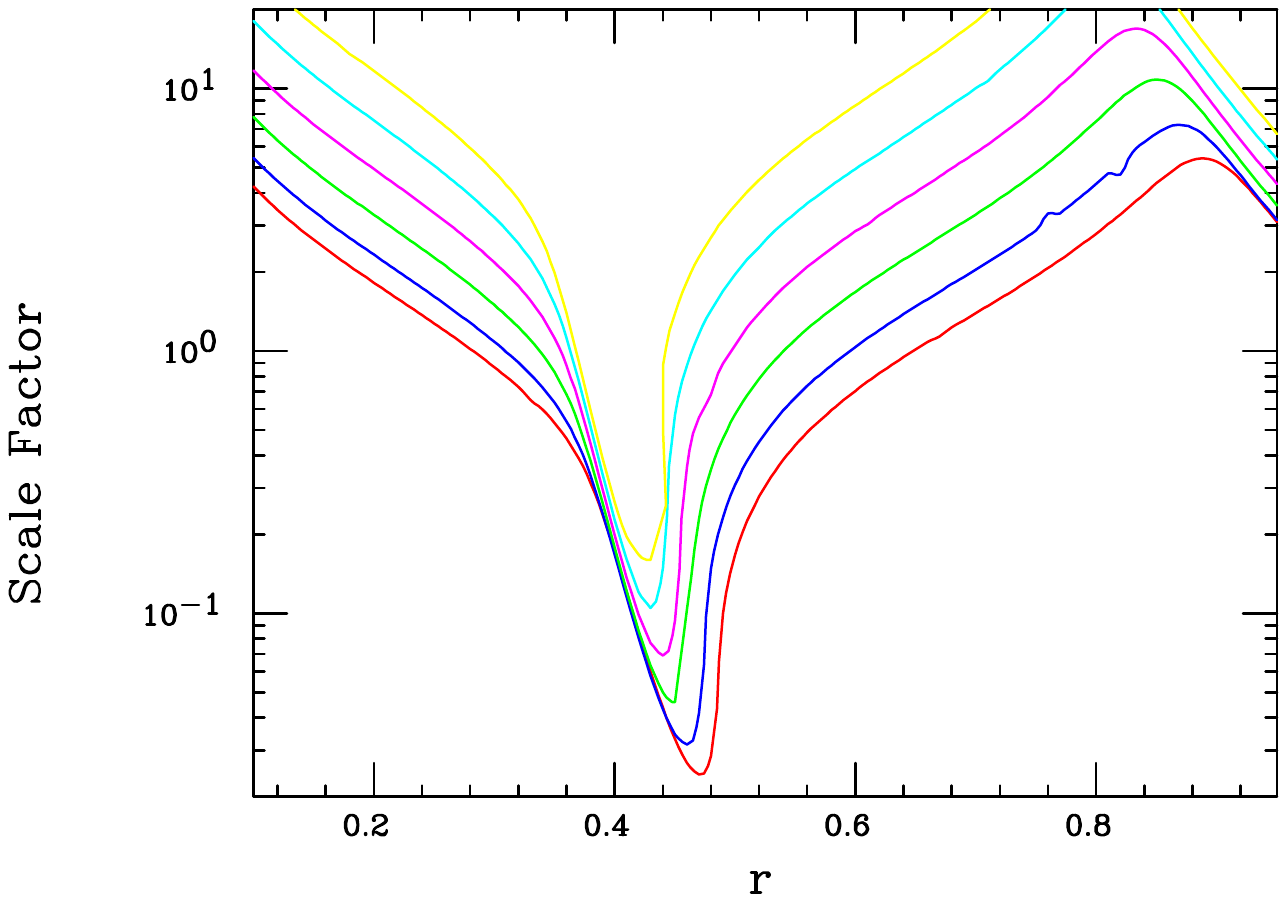}}
\vspace*{-1.3cm}
\caption{Same as the previous Figure but now for (Top) $\lambda_R^2=6$ with $t_I=1.0$ and (Bottom) $\lambda_R^2=7$ with $t_I=0.90$.}
\label{fig3}
\end{figure}

\begin{figure}[htbp]
\centerline{\includegraphics[width=5.0in,angle=0]{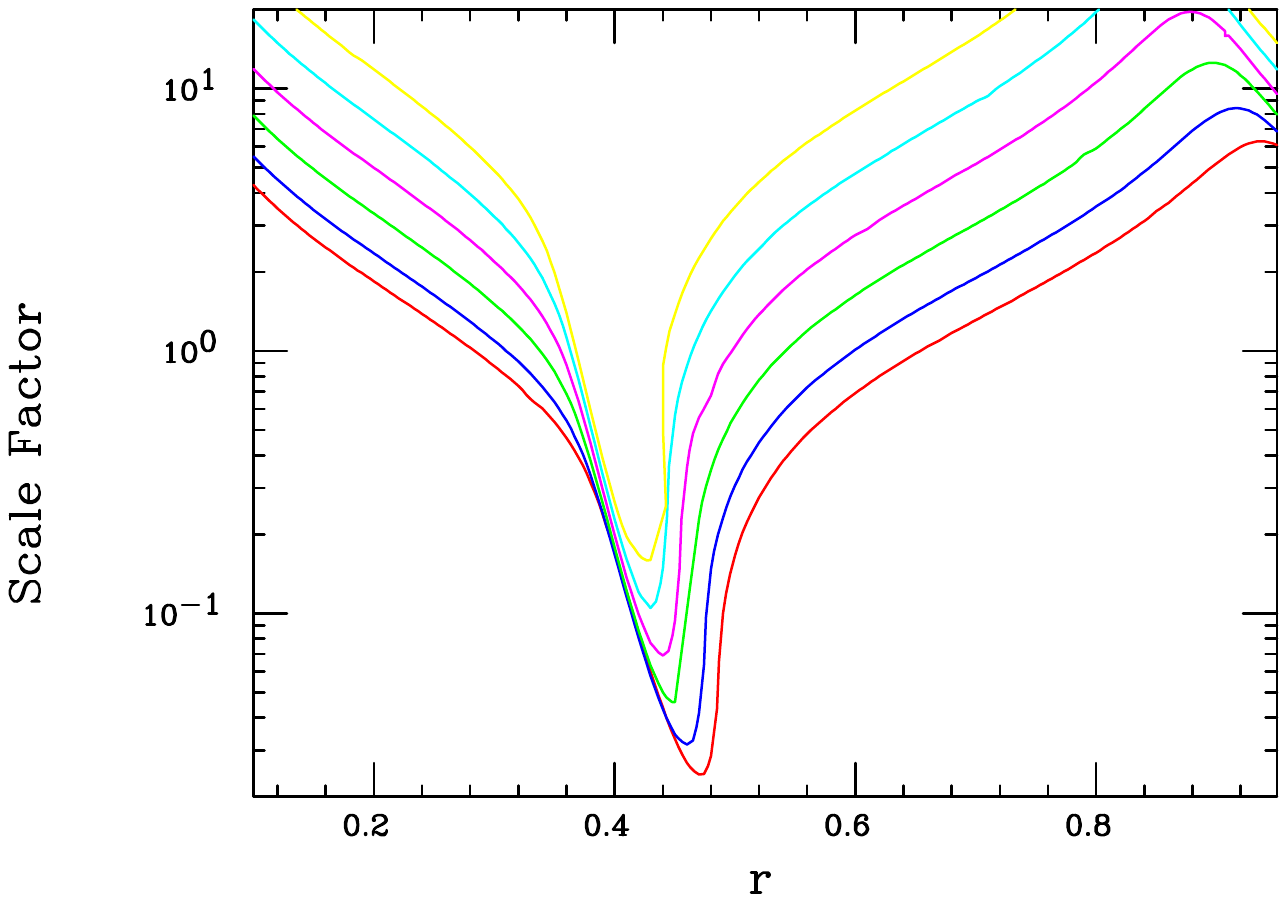}}
\vspace*{-0.8cm}
\centerline{\includegraphics[width=5.0in,angle=0]{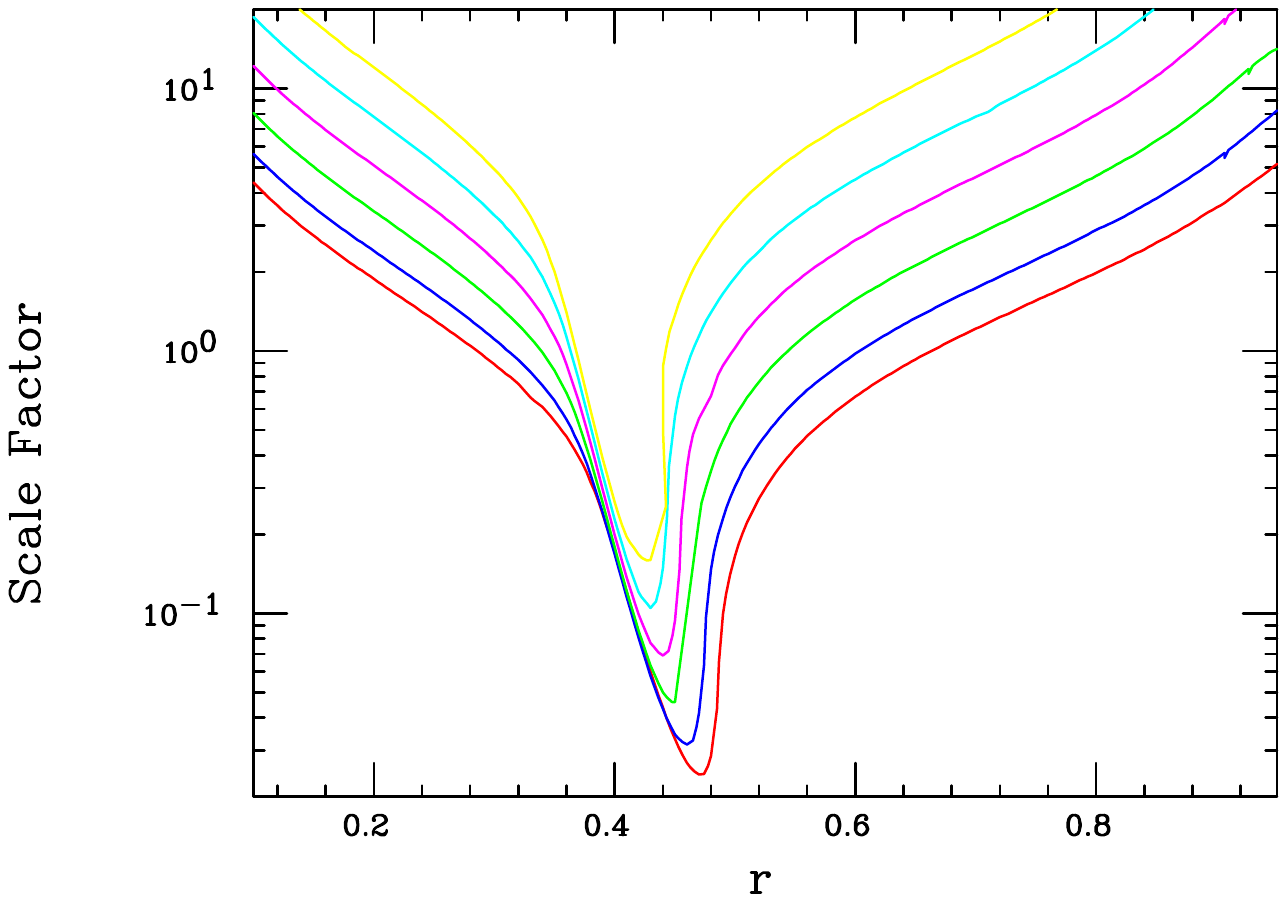}}
\vspace*{-1.3cm}
\caption{(Same as the previous Figure but now for (Top) $\lambda_R^2=8$ with $t_I=0.85$ and (Bottom) $\lambda_R^2=9$ with $t_I=0.80$.}
\label{fig4}
\end{figure}

In Fig.~\ref{fig2} we can examine what happens as one slowly increase the value of $\lambda_R^2$ to 4(5) in the upper (lower) panel; significant changes from the case of $\lambda_R^2=3$ 
are clearly observed. When $\lambda_R^2=4$, the choice $t_I=1.5$ still leads to a constructive interference between the two resonances but they are now spaced further apart with at least some 
of the greatest activity appearing beyond the limited $r\leq 0.9$ range that we consider here. 
This leads to a modest increase in the value of ${\cal F}$ required to reproduce the observed DM relic density in the region between the two resonances which, however, remains relatively preferred 
especially for smaller values of the parameter $\delta$. Once we go beyond $\lambda_R^2=4$, the value of ${\cal R}$ is found to change sign for our chosen benchmark values of $t_I$, thus leading 
to a destructive interference between the $Z_{1,2}$ exchanges. This effect is not immediately obvious in the case of $\lambda_R^2=5$ as the $Z_{1,2}$ resonances are still spaced rather closely 
so that the cross section does not have enough room to dive to 
particularly small values and correspondingly larger values of ${\cal F}$. This increase in the required value of ${\cal F}$ to reproduce the observed relic density when going from $\lambda_R^2=4$ 
to $\lambda_R^2=5$ does not appear to be significantly larger that than seen when going from the case of $\lambda_R^2=3$ to that of $\lambda_R^2=4$. Here we see that in this interference 
regime, values of $\delta \lsim 0.15-0.2$ will still remain in the preferred range. 
When $\lambda_R^2$ reaches values $\gsim 6$, we see in Figs.~\ref{fig3} and ~\ref{fig4} that the now stronger effect of the destructive interference results in pushing the ${\cal F}$ 
needed to reproduce the DM density significantly upward to larger values. When $\lambda_R^2$ increases even further to a value of 8 or 9, we start to see in Fig.~\ref{fig4} many of the aspects 
associated with the single exchange limiting case, as is shown in Fig.~\ref{fig1}, beginning to be reproduced and the majority the effects of the destructive interference now occurs outside of the 
restricted range of $r \leq 0.9$. Obviously, once $\lambda_R^2$ exceeds these values, one will obtain results which will appear to be very much like those seen in Fig.~\ref{fig1}. 

Clearly, we've learned from this analysis that if $\lambda_R^2$ is sufficiently small so that constructive interference occurs in the the region above the $Z_1$ or that destructive interference 
remains relatively ineffective then that range of the DM mass now becomes much more favorable in comparison to what occurs in the case when there is only a single DP exchange. Of course, in 
the present case, we've relied on a specific setup to demonstrate this result and we are to be reminded that other choices of the dark Higgs fields would re-weight the couplings of the $Z_{1,2}$ 
gauge bosons in a somewhat different manner; however, our expectation is that what we have found above would be a rather qualitative general result although the specific details would be model 
dependent. 

Finally, we need to return to the question of how the $\chi_2$ decays; since $r\delta \leq 0.27 <1$ in the parameter range of interest to us here, the decay $\chi_2 \to \chi_1 Z_1$ cannot occur on-shell. 
Also, since 
the light Higgs mass eigenstate fields, $h_a$, we encountered above only couple diagonally to the $\chi$'s, there are no allowed decays of the type $\chi_2 \to \chi_1h_a$.  Instead, one needs to 
focus on the 3-body decay process via the virtual exchange of the two light gauge bosons, \eg, $\chi_2 \to \chi_1 Z_{1,2}^* ,Z_{1,2}^* \to e^+e^-$. In the case of such a decay being mediated by 
a single DP, $V$, with coupling $g_D$, this rate is well-known\cite{Garcia:2024uwf} to be approximately given by the expression (in the limit where $(r\delta)^2 <<1$ is assumed so that any energy 
dependence in the gauge boson propagators can be ignored) 
\begin{equation}
\Gamma_{3-body}\simeq \frac{4m_2^5}{15\pi m_V^4}\alpha_D \alpha \epsilon^2 \delta^5 \,,  
\end{equation}
whereas in the present setup, both of the two $Z_i$ exchanges will simultaneously contribute to this process. In such a case, one would then 
expect, in the limit where this approximation is valid, that the decay rate would be roughly scaled by an overall factor of 
\begin{equation}
R_2=\Bigg(1+f\frac{{\cal R}}{\lambda_R^2}\Bigg)^2\,,  
\end{equation}
which may be quite substantial until larger $\lambda_R^2$ values are reached. Here, $f=f(r\delta,\lambda_R^2)$, is a correction factor, which we find differing from unity by at most a few 
percent, accounting for finite $(r\delta)^2$ effects in our parameter range of interest. The values of this ratio, $R_2$, for our set of benchmark points is also shown in Table~\ref{qtab} where we see that 
both destructive and constrictive interference effects are possible depending upon our location in parameter space; note that for finite $r\delta$, this effect can be only slightly degraded.  Given the 
value of $R_2$, we can then express the corresponding un-boosted decay length for this $\chi_2$ decay as
\begin{equation}
c\tau \simeq 4.8\times 10^{6} ~{\rm cm}~ \Bigg[\Big(\frac{M_{Z_1}}{100 ~{\rm MeV}}\Big)~\Big[\frac{g_{eff}\epsilon}{10^{-4}}\Big]^2~\Big(\frac{I}{10^{-7}}\cdot R_2\Big)\Bigg]^{-1}\,.  
\end{equation}

\begin{figure}[htbp]
\centerline{\includegraphics[width=5.0in,angle=0]{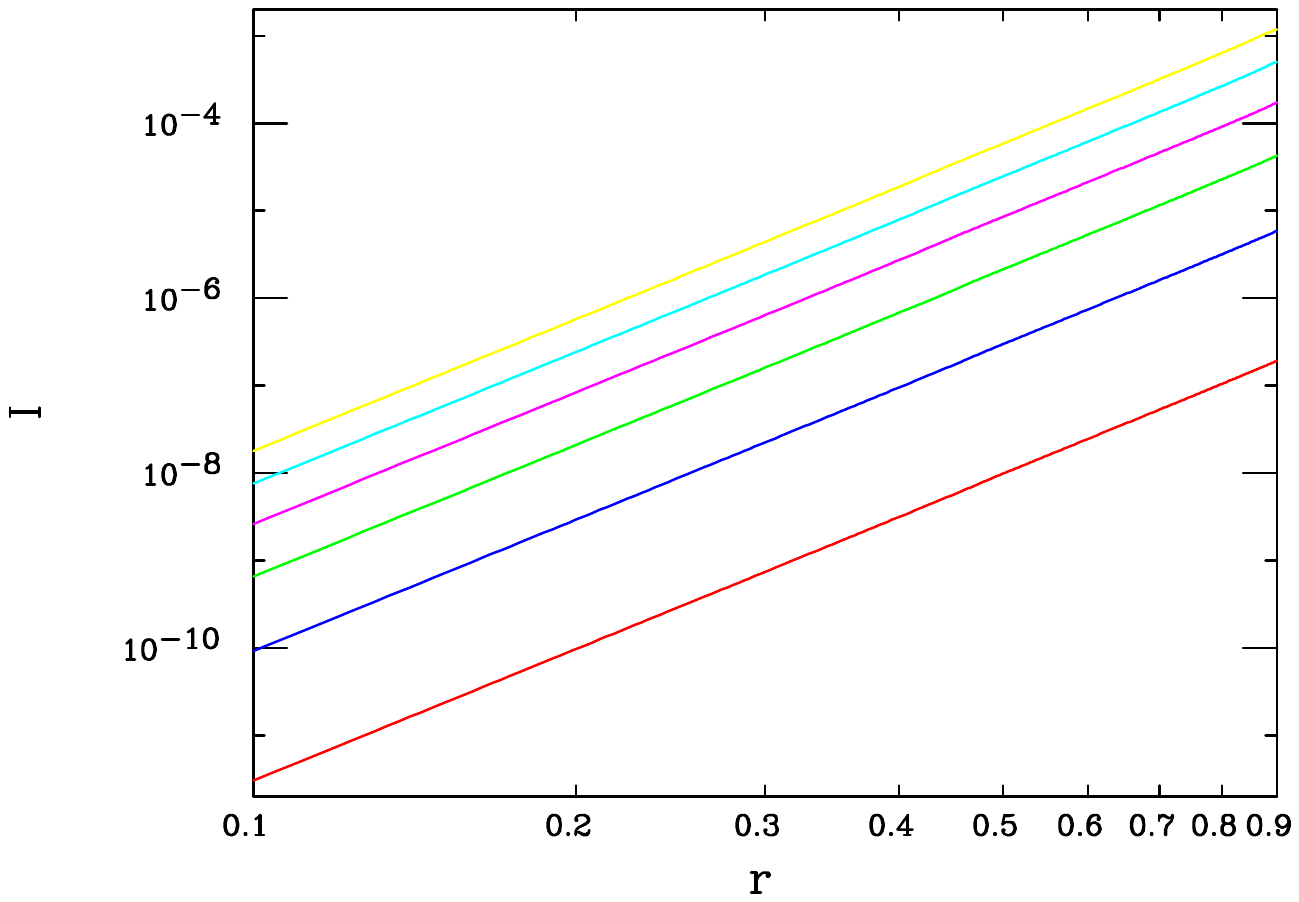}}
\vspace*{-1.3cm}
\caption{The phase space integral, $I$, as a function of $r$ for different values of $\delta$ as described in the text. The curves from bottom to top correspond to different choices of $\delta$ from 
0.05 to 0.30 in steps of 0.05, respectively.}
\label{figz}
\end{figure}

Fig.~\ref{figz} shows the numerical results for the 3-body phase space integral, $I$, appearing in the formula above as a function of $r$ for different values of $\delta$ in the limit that $m_e^2 \to 0$. 
Here we can see, amongst other things, the approximate validity of the $\sim (r\delta)^5$ scaling that we might expect from the approximate decay width formula above and that $I$ takes on 
a very wide range of values, $\sim 10^{-12}-10^{-3}$, as both $r$ and $\delta$ are varied over their allowed ranges. The anticipated parameter scaling can be seen in both the approximate power 
law behavior with $r$ as well as the approximately 
constant spacing between the curves corresponding to different values of $\delta$. Note that for the parameter ranges that we consider here, $I$ is such that one obtain lifetimes values 
corresponding to macroscopic un-boosted decay lengths so that $\chi_2$ will appear as either just MET or MET + a displaced $e^+e^-$ pair in a detector depending upon how it is produced even 
when constructive interference between the two $Z_i$ is present and interference effects are relevant. These are important results to have in hand before the considerations of the production of 
the new heavy states in our setup to be discussed in the next Section.

\section{Phenomenological Implications: Present and Future High Energy Colliders}

As discussed earlier, unlike in the PM models previously considered, the high energy collider phenomenology of the present model is somewhat limited. The reasons for this are twofold: ($i$) First, 
the two PM fermions in the current setup are paired alone together within their own $SU(2)_I$ representation which, in the absence of very small mixing, is completely isolated from the SM. This means 
that the non-hermitian $W_I^\pm$ 
gauge bosons do not couple directly to any of the SM fields. ($ii$) Both of the light hermitian gauge bosons, $Z_{1,2}$, only couple to the SM via suppressed KM interactions.  Thus, \eg, $W_I$ pair 
production via a TeV-scale, and potentially resonant,  $Z_I$ exchange, where the $Z_I$ previously had unsuppressed couplings to the SM quarks or leptons in the initial state, is now absent. 
This is unfortunate as new neutral heavy gauge bosons can provide an important window onto new physics and are relatively accessible\cite{ATLAS:2019erb,CMS:2021ctt,Helsens:2019bfw}. A reaction 
similar to this via {\it light} $Z_i$ exchanges, but which is thus never resonant, {\it does} occur but its rate is $\epsilon^2$ suppressed as we'll discuss briefly below. The other possible channel, 
$W_I$ associated production with a heavy PM fermion via, \eg, $gq$-fusion, is also absent here since, again, the PM does not share any representation with the SM fermions.  On top of this, all of the 
new Higgs fields needed to break the various non-SM gauge symmetries are dark and so will not couple directly to any of the SM particles in the absence of mixing which for the most part  
cannot be too large due to the experimental constraints on the H(125) properties.

\begin{figure}[htbp]
\centerline{\includegraphics[width=5.0in,angle=0]{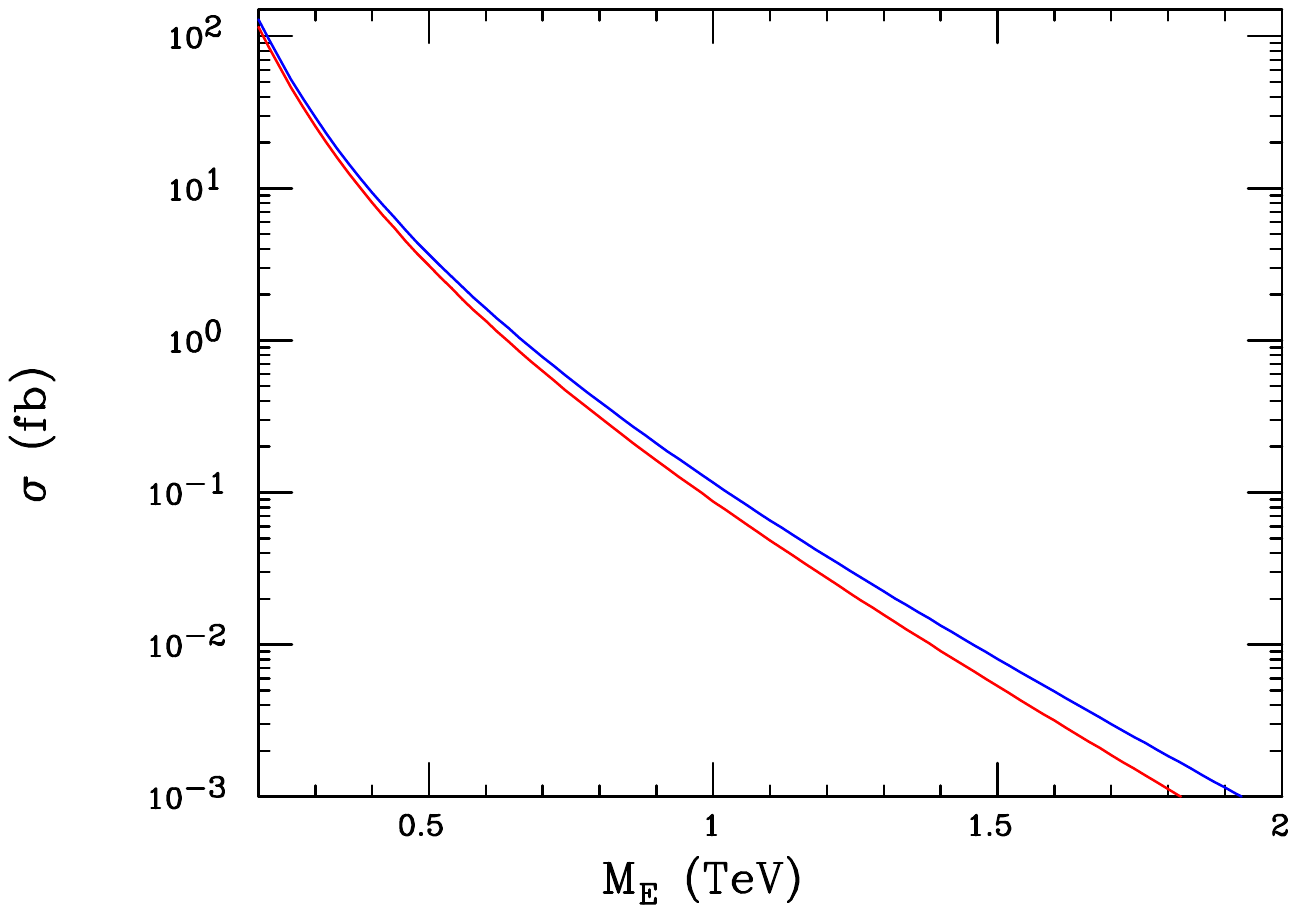}}
\vspace*{-0.8cm}
\centerline{\includegraphics[width=5.0in,angle=0]{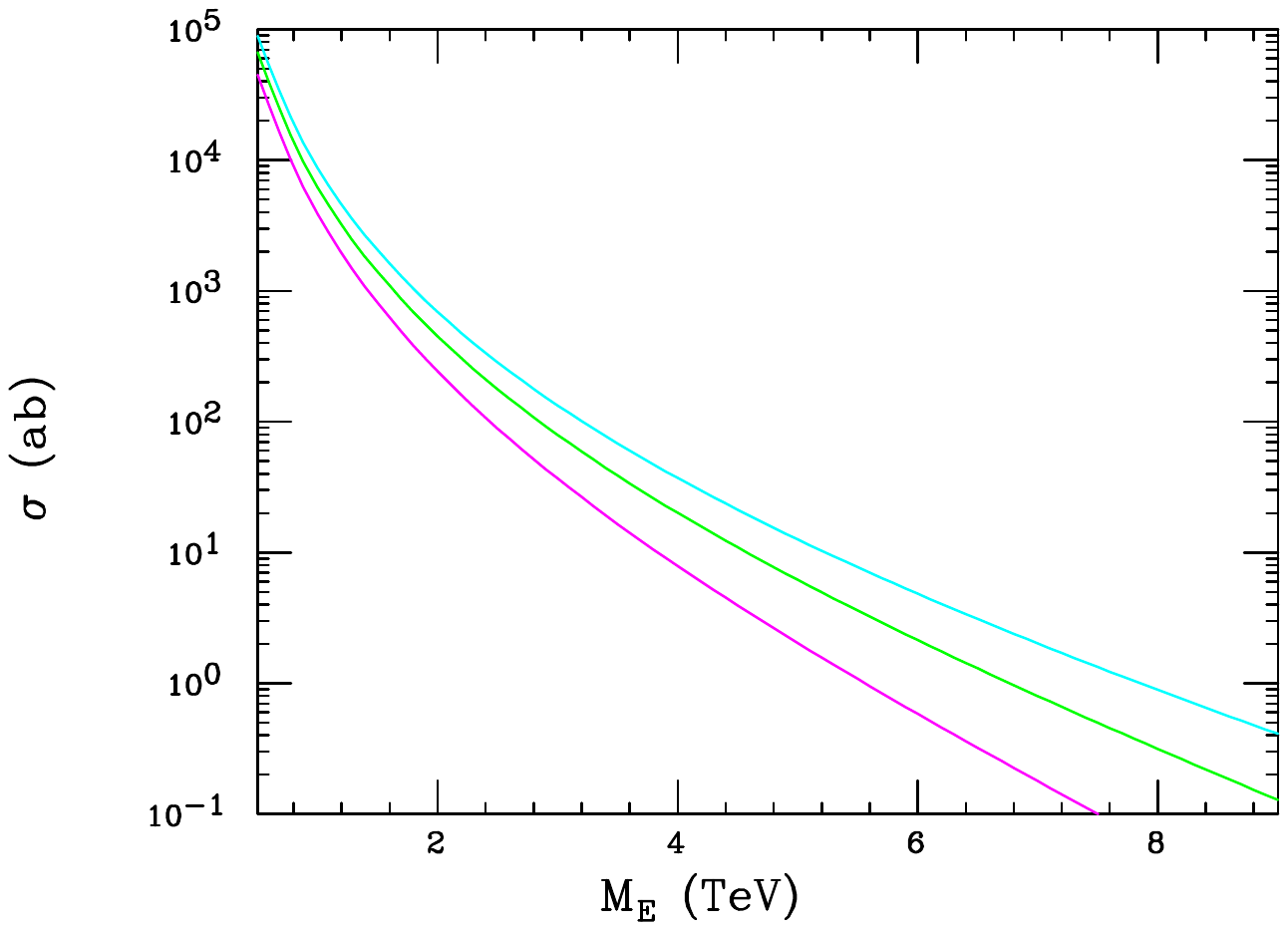}}
\vspace*{-1.3cm}
\caption{Pair production cross sections for the SM isosinglet, charged lepton-like PM fermion at hadron colliders via $s$-channel $\gamma,Z$ exchange: (Top) at the $\sqrt s=13(14)$ TeV 
LHC represented by the red (blue) curve, respectively, and (Bottom) at the $\sqrt s=100(80,60)$ TeV FCC-hh, represented by the cyan (green, magenta) curve, respectively.}
\label{fig5}
\end{figure}

The PM fermions themselves, here assumed to be $SU(2)_L$ isosinglet charged leptons for concreteness, can only be produced by the usual electroweak interactions via $\gamma,$Z exchanges, 
at relatively suppressed rates and so it 
is more than likely that only the lighter of these two states, $E_2$, might be accessible to the LHC and possibly at future colliders as well. In fact, as we'll see in the Figures discussed below, 
increasing the PM mass by $\sim 50\%$ results in a reduction of the production cross section by more than an order of magnitude. As was discussed above, the mixing of this state with a SM 
isosinglet, RH-lepton, denoted generically as $e_R$ here, leads to the dominant decays $E_2\to e Z_{1,2}$. If the $Z_i$ decay invisibly or are are sufficiently boosted, the pair production of $E_2$'s 
will appear as opposite-sign dileptons plus MET (qualitatively similar to that for isosinglet sleptons), which suffers a significant SM backgrounds from $W^+W^-$ pair production with both $W$'s 
simultaneously decaying into same flavor charged leptons plus neutrinos. The pair production cross sections for these states at the LHC and at FCC-hh with various center of mass energies 
are shown for reference in the top and bottom panels of 
Fig.~\ref{fig5}, respectively, and are smaller than those for all other possible charged VL fermionic PM states implying that only a relatively poor PM search reach is obtainable. 

Using $139$ fb$^{-1}$ of integrated luminosity at the 13 TeV LHC, the authors of Ref.\cite{Guedes:2021oqx} have recast the then existing ATLAS and CMS SUSY searches to obtain limits on the pair 
production of isosinglet vector-like leptons decaying with just this pattern of interest here, \ie, $E^+E^- \to e^+e^- +$MET, obtaining a lower bound on the mass of such a particle that we would 
interpret here as being $M_{E_2}>895$ GeV. These same authors have also extrapolated these results to obtain possible exclusion (not discovery!) reaches at the HL-LHC with an luminosity of 
3 ab$^{-1}$ of 1.45 TeV (but still assuming that $\sqrt s=13$ TeV) and at the $\sqrt s=100$ TeV FCC-hh, also with an assumed integrated luminosity of 3 ab$^{-1}$, of 3.33 TeV.  An increase in 
the integrated luminosity by an order of magnitude may be able to push this mass reach of the FCC-hh up into the neighborhood of $\simeq 4.5$ TeV whereas a reduction of $\sqrt s$ to 60(80) TeV 
may result in a decreased reach of roughly $\simeq 30(13)\%$ or so as indicated by this Figure.

Of course if we had instead chosen the PM fields to be color triplets, as was briefly mentioned above, \ie, $U_i$ and/or $D_i$, then these fields would be pair-produced via QCD interactions and the 
relevant final state signature would then be 2 jets + MET, possibly with flavor tagging, leading to somewhat greater search reaches at hadron colliders by factors of roughly 
$\simeq 1.5-3$\cite{Rizzo:2022qan} depending upon the assumed collider energy and luminosity, \eg, $\simeq 10$ TeV or more at the $\sqrt s=100$ TeV FCC-hh with 30 ab$^{-1}$ of 
integrated luminosity.  

\begin{figure}[htbp]
\centerline{\includegraphics[width=5.0in,angle=0]{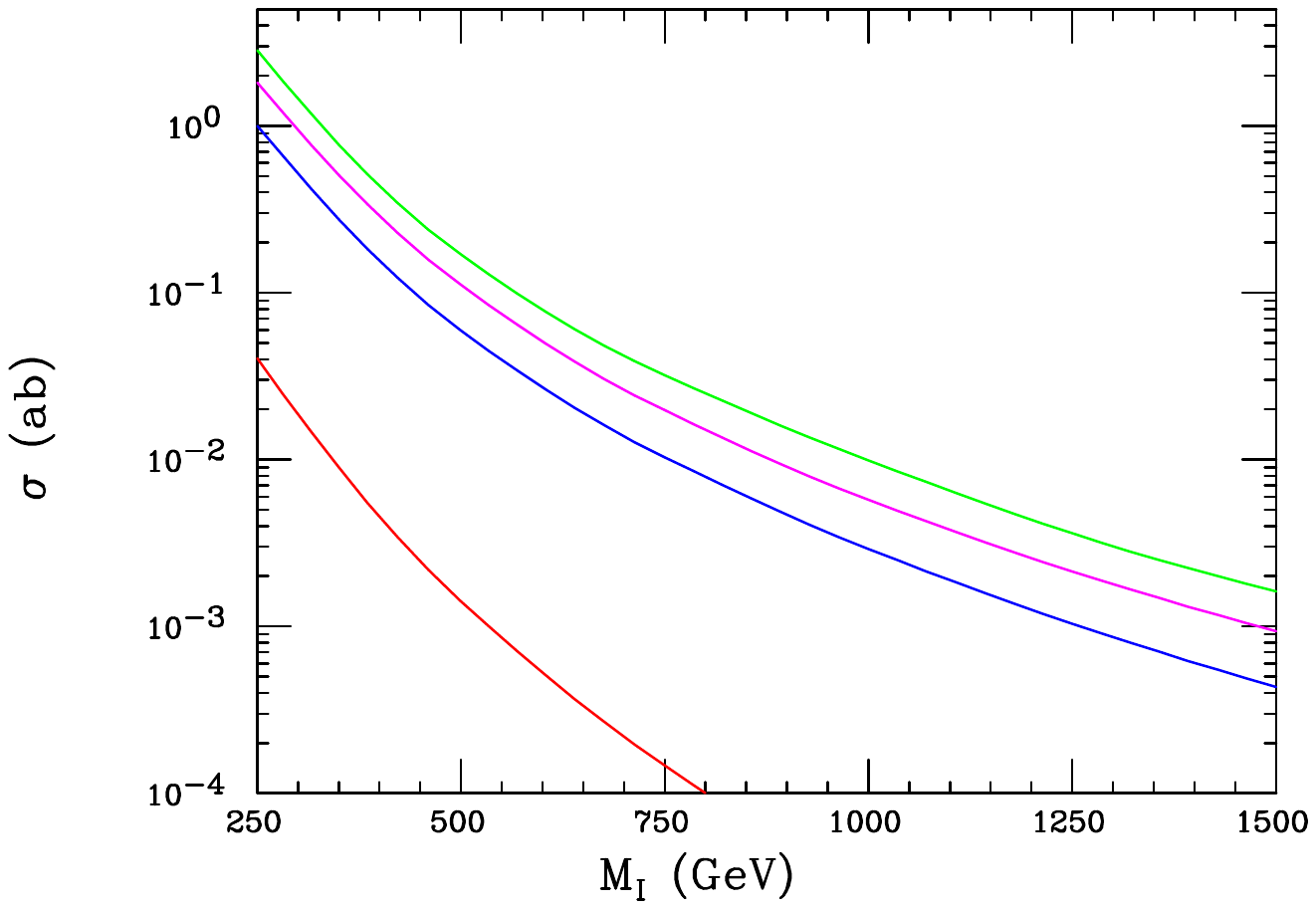}}
\vspace*{-0.8cm}
\centerline{\includegraphics[width=5.0in,angle=0]{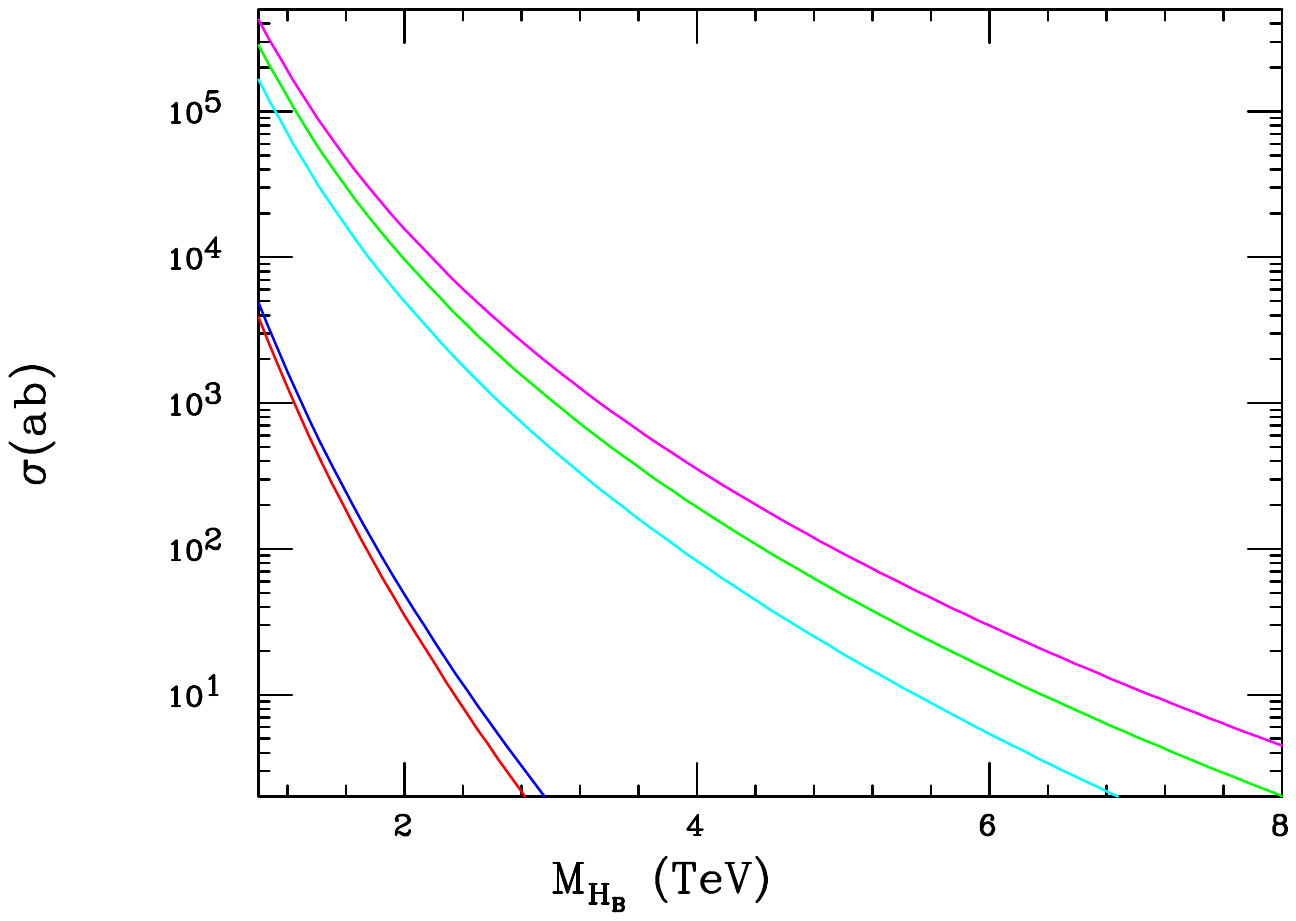}}
\vspace*{-1.3cm}
\caption{(Top) $W_IW_I^\dagger$ production cross section as a function of its mass at the $\sqrt s=14$ TeV LHC and the $\sqrt s=100(80,60)$ TeV FCC-hh corresponding to the 
green(magenta, blue) curves , respectively, assuming that  
$g_I\epsilon=10^{-4}$, as discussed in the text. (Bottom) $H_B$ production cross section as a function of its mass via $gg$-fusion mediated by top quark loops assuming that 
$s^2_\lambda=0.025$ for purposes of illustration as discussed in the text. The red (blue) curve is for the $\sqrt s=13(14)$ TeV LHC whereas the magenta (green, cyan) curve corresponds 
to the $\sqrt s=100(80,60)$ TeV FCC-hh, respectively.}
\label{fig6}
\end{figure}

As noted above, the heavy $W_I$'s may be pair produced in a non-resonant manner from SM initial states via the $s$-channel exchange of the light $Z_{1,2}$ gauge bosons but with rates which 
are necessarily $\epsilon^2$ suppressed. In fact, the mass of the $W_I$ is so large compared to those of the $Z_i$ that we can obtain the desired cross section simply from that due to the exchange 
of a single, massless 
$W_{3I}$ gauge boson state (since the longitudinal modes don't couple to the initial state), thus removing much of the apparent model dependency. Except for the $W_I$ mass itself and the 
numerical value of the product, $g_I\epsilon$, the (subprocess) differential 
cross section for $\bar f f\to W_IW_I^\dagger$ is completely determined and is provided by the PDG\cite{ParticleDataGroup:2024cfk} after some slight modifications, \ie,  
\begin{equation}
\frac{d \sigma}{dz}= \frac{\alpha}{8N_cs} ~(g_I\epsilon)^2 ~Q_f^2~\beta_w~\Bigg[\Big(\frac{tu}{M_I^4}-1\Big)~\Big(\frac{1}{4}-\frac{M_I^2}{s}+3\frac{M_I^4}{s^2}\Big)+\frac{s}{M_I^2}-4\Bigg]\,,
\end{equation}
where $t,u$ are the usual Mandelstam variables which are explicitly given by just the familiar expressions, \ie,
\begin{equation}
t,u=M_I^2-\frac{1}{2}s(1\mp \beta_wz)\,,
\end{equation}
with $\beta_w^2=1-4M_I^2/s$, $z=\cos \theta$ being the $W_I$ scattering angle, $N_c=1,3$ is the color factor and $Q_f$ is the electric charge of the initial state fermion. Assuming for 
demonstration purposes 
the $g_I\epsilon=10^{-4}$, similar to the discussion above, integrating the above differential cross section over $z$ and the relevant parton densities, we obtain the numerical result for this cross 
section at the LHC and FCC-hh for different assumed center of 
mass energies as shown in the upper panel of Fig.~\ref{fig6}. Here we see that, independently of how these states might decay, the pair production cross section for $W_I$ is far too small to 
be observable at the HL-LHC even if $M_I$ were to be as low as only 250 GeV. The $\sqrt s=100$ TeV FCC-hh, however, fares somewhat better but even there, with an integrated luminosity of 
30 ab$^{-1}$, doesn't provide a mass reach beyond, perhaps, $\sim 500$ GeV, a value which might be further eroded if the center of mass energy is lowered and depending upon how the 
$W_I$'s decay itself might be tagged. It seems more than likely that any direct access to the physics of the $W_I$ must be obtained by other means than via this process.

Although the SM Higgs is highly constrained not to mix with light dark Higgs that may yield H(125) decays to invisible final states at too large of a rate as was discussed above, the mixing of the usual 
Higgs with the {\it heavier} dark Higgs living at the $\sim v_T$ scale is a bit less constrained. Those Higgs scalars, to lowest order, will only couple to the heavy dark sector fields, $W_I, E_{1,2}$, \etc, 
which are all kinematically inaccessible to H(125). Thus we see that their mixing with the SM Higgs will essentially only lead to an overall reduction in the H(125) total width and will leave all of the 
branching fractions unaltered, something which is a bit more subtle to detect by the LHC experiments. The real, $T_{3I}=Y_I=0$ scalar, $\Sigma^0$, is an example of such a field which may experience 
a small $\sim v_{SM}/v_T$-scale mixing with the SM Higgs and may be able to be produced at a visible rate via this mixing. We recall that the other scalars in $\Sigma$ carrying $T_{3I}=\pm 1$,  
in the ratio of large vevs limit, no longer remain in the physical spectrum as they are essentially the Goldstone bosons which are eaten by the $W_I^\pm$ when they obtain mass. 

If we denote the heavy scalar mass eigenstate (assuming only mixing with the SM Higgs for simplicity in obvious notation) by $H_B=c_\lambda\Sigma^0+s_\lambda h_{SM}$, 
(with $h=H(125)$ then being identified with the lighter, orthogonal combination, $H_A$) we can employ, \eg, the process $gg\to H_B$ via the top quark loop (assuming that any heavy colored PM 
fields are absent in the present setup) to estimate this cross section at hadron 
colliders; this is shown in the lower panel of Fig.~\ref{fig6} for the $\sqrt s=13,14$ LHC and also for the FCC-hh with $\sqrt s=60, 80$, or $100$ TeV.  Here for purposes of demonstration 
we have assumed a $2.5\%$ {\it overall} suppression of the H(125) couplings (\ie, $s_\lambda^2=0.025$) relative to the SM which still remains relatively safe within the experimentally allowed 
region\cite{Tackmann:2023wfs,Moriond25,Heo:2024cif,Heo:2025vkz} from current fits to the H(125) coupling data. Note that unitarity constraints on electroweak singlet heavy Higgs fields from 
SM $WW$ scattering further require that\cite{Kang:2013zba} $M_{H_B}|s_\lambda | \lsim 1$ TeV which in our numerical example roughly corresponds to the bound $M_{H_B}\lsim 6.3$ TeV. 
Note further that we've ignored the possible corresponding production of $H_B$  employing the $WW$ fusion process as it 
is quite sensitive to both $M_{H_B}$ as well as the specific value of $s_\lambda^2$ for just this reason. Thus the cross section we display here is actually the {\it minimum} result for this chosen 
value of $s_\lambda^2$. Of course this situation may change significantly if heavy colored PM fields coupling to $H_B$ are also/were instead present in an augmented version of this setup. 
This would be the case if we had chosen color triplet instead of color singlet PM fields as has been discussed earlier. 

Since $H_B \simeq \Sigma^0$ couples to pairs of $W_I$'s (as well as the $E_i$'s as noted), their production via this intermediary is far likely dominant, especially so if the 
$H_B\to W_IW_I^\dagger$ decay can occur on-shell as it can then easily also be the dominant decay mode.  If these heavy modes are not kinematically accessible on-shell, then decays to SM final 
states, in particular $W^+W^-$ due to the rather strong coupling that yields the unitarity bound mentioned above, may become of significant relevance and will dominate. 

One can imagine other processes that can be induced via the mixing between the SM Higgs and $\Sigma^0$ at the loop-level, \eg, $h \to \gamma +Z_i$, as generated by the lepton-like PM fields. 
However, the rate for such a process is not only suppressed by an overall loop factor, $\sim \alpha^3/\pi^2$ and by small mixings, $s_\lambda^2 \sim 0.025$, but also by the small mass ratios,  \eg, 
$m_h^2/M_I^2 \lsim 10^{-2}$. A rough estimate of the sum of partial widths for these processes is given to leading order in the small mass ratios, \ie, $M_{Z_i}^2<< m_h^2<< M_{E_i}^2$, by 
\begin{equation}
\sum _i ~ \Gamma(h \to \gamma +Z_i) \simeq \frac{m_h \alpha^3}{288 \pi^2} ~s_\lambda^2~\Big(\frac{\alpha_I}{\alpha}\Big)^2~\Big(\frac{m_h}{M_I}\Big)^2 ~\Bigg[\frac{M_{E_1}^2-M_{E_2}^2}
{M_{E_1}M_{E_2}}\Bigg]^2\,,
\end{equation}
where, employing typical values of the various parameters and using the expected SM Higgs total width of $\simeq 4.1$ MeV, leads to a branching fraction of $B\sim 10^{-9}-10^{-8}$ which is 
quite inaccessible. A rather similar result, up to factors of a few, would also be found to hold if color triplet PM fields as discussed above had instead been chosen.

As noted above, in the limit that $v_T^2 >> v_{SM}^2>>v_{1,2}^2$, the only other (beyond the SM Higgs and $\Sigma^0$) remaining physical Higgs field degrees of freedom that are {\it not} eaten as 
Goldstone bosons are those solely arising from the dark $SU(2)_I$ doublet and triplet scalars, $H_{1,2}$, respectively, none of which carry SM quantum numbers. Needless to say, this makes 
studying any of the physics associated with them rather difficult to explore directly. This, however, also happens in a similar fashion in the more conventional `SM-like' symmetry breaking pattern 
as there the heavy $Z_I$ eats an additional Goldstone boson occurring in the $SU(2)_I$-breaking Higgs doublet employed therein. Thus we may want to further explore indirect collider signatures 
for the new physics induced by the setup above. 

As a step in this direction, we consider the pair production of the light dark gauge bosons and/or light dark Higgs at future $e^+e^-$ colliders paralleling the analysis procedure as employed in 
Ref.\cite{Rizzo:2023djp}. To be specific, we consider the ten (sub-)processes $e^+e^- \to Z_iZ_j, h_ah_b,Z_ih_a$, with $i=1,2;~a=1,2$, all of which we will assume produce an invisible final state 
via their decays down to DM. These processes dominantly arise from the $t-,u-$channel exchanges of the lepton-like PM fermions which are induced by the couplings resulting from the
required $E_2-e_R$ mixing, described in Eq.(14), via the Yukawa coupling $y_{mix}$. In what follows we will assume that $y_{mix}$ takes the value of unity as our results are then easily scalable.  Since these particles are all invisible, we perform the usual trick of tagging this final state via an initial state photon (\ie, ISR) and so we are actually looking for the well-known 
$e^+e^- \to \gamma +$`nothing' production mode. Noting that the masses of the $Z_i$ and $h_a$ are all far smaller than the nominal $\sqrt s=1$ TeV value that we will consider below, we can 
treat them as (essentially) massless and employ the Goldstone Boson Equivalence Theorem \cite{GBET} to perform our cross section calculation so that their individual contributions can all be easily 
summed together. This greatly simplifies the result as individual dependencies on the various mixing angles will vanish when this sum is performed so that only the overall scale set by $ y_{mix}$ 
remains an unknown. To calculate the relevant cross section, 
as in earlier work, we will make use of an improved radiator function, ${\cal R}_\gamma$, as discussed in, \eg, Ref.\cite{Chu:2018qrm,Choi:2015zka} so that we can write the physical 
$e^+e^-\to \gamma$ plus light dark boson cross section in the form
\begin{equation}
\frac{d\sigma}{dx_\gamma~d\cos \theta_\gamma}=\sigma_D(s\to \hat s)~{\cal R}_\gamma(x_\gamma,\theta_\gamma)\,,  
\end{equation}
where we define the parameter $a=1+2M_E^2/s$ so that $\sigma_D(s)$ is just given by (after summing over the various final states with their proper mixing angles and weighting factors), 
\begin{equation}
\sigma_D=\frac{y_{mix}^4}{64\pi s} ~\Bigg[\frac{(3a^2-1)}{a}~\tanh ^{-1} \Big(\frac{1}{a}\Big)-3\Bigg]\,.  
\end{equation}
Note that this cross section rapidly falls rapidly as $\sim (\sqrt s/M_E)^8 \to 0$ for small center of mass energies thus requiring us to look toward measurements at higher energy colliders and not 
at experiments such as, \eg, BELLE-II, with small $\sqrt s$.  In the expression above, $x_\gamma=2E_\gamma/\sqrt s$ is the scaled energy of the ISR photon so 
that $\hat s=(1-x_\gamma)s$ is the $e^+e^-\to$ light dark boson pair subprocess center of mass energy squared. Here too, $\theta_\gamma$ is the photon's scattering angle as measured 
from the $e^-$ beam direction in the center of mass frame. The radiator function that we will again use, in the limit of vanishing electron mass, is then just given by\cite{Chu:2018qrm,Choi:2015zka}
\begin{equation}
{\cal R}_\gamma (x_\gamma,\theta_\gamma)=\frac{\alpha_{em}}{\pi}~\frac{1}{x_\gamma}~\Bigg[\frac{1+(1-x_\gamma)^2}{1-\cos^2 \theta_\gamma}-\frac{x_\gamma^2}{2}\Bigg]\,,  
\end{equation}
so that the integration over $z=\cos \theta_\gamma$ is trivial once the relevant experimental detector angular acceptance cuts, $-z_0\leq z\leq z_0$, are imposed. Now as is well-known, this 
final state suffers from a huge SM background coming from the sub-process $e^+e^-\to $neutrino pairs that arises from both $s$-channel $Z$ and, dominantly, $t-$channel $W$ exchanges; we 
will make use of the analytical results for this SM cross section that are found in Refs.\cite{Hirsch:2002uv,Barranco:2007ej,Berezhiani:2001rs,Escrihuela:2019mot} to estimate this background. 
We also recall that this SM contribution arises 
almost completely from the left-handed electron polarization state whereas the new physics signal we seek arises instead from {\it right-handed} electrons. This then behooves us to focus on
linear colliders, such as the ILC, C$^3$ or CLIC \cite{ILC:2007oiw,ILCInternationalDevelopmentTeam:2022izu,Aicheler:2012bya,Bai:2021rdg,LinearColliderVision:2025hlt} as they will allow for 
highly polarized $e^\pm$ beams, \eg, $|P_{e^-}|=0.8$ and $|P_{e^+}|=0.3$, to reduce this sizable SM background. Making a judicious choice of these polarizations, a purely LH-coupling, \eg, 
$e_L^-e_R^+$, as that induced by the SM $W$ coupling{\footnote {The much smaller $Z$ exchange process contributes roughly equally to both left-and right-handed polarizations.}} can then 
be suppressed by a factor of $(1-0.8)(1-0.3)=0.14$ while with this same choice a purely RH-coupling, \eg, $e_R^-e_L^+$, as occurs 
here in the PM-electron interaction, can  be simultaneously enhanced by a factor of $(1+0.8)(1+0.3)=2.34$; this has been noted by the authors of numerous other BSM studies employing this particular 
final state\cite{Chae:2012bq,Kalinowski:2021tyr,Khalil:2021afa,Black:2022qlg,Dreiner:2012xm}. 

\begin{figure}[htbp]
\centerline{\includegraphics[width=5.0in,angle=0]{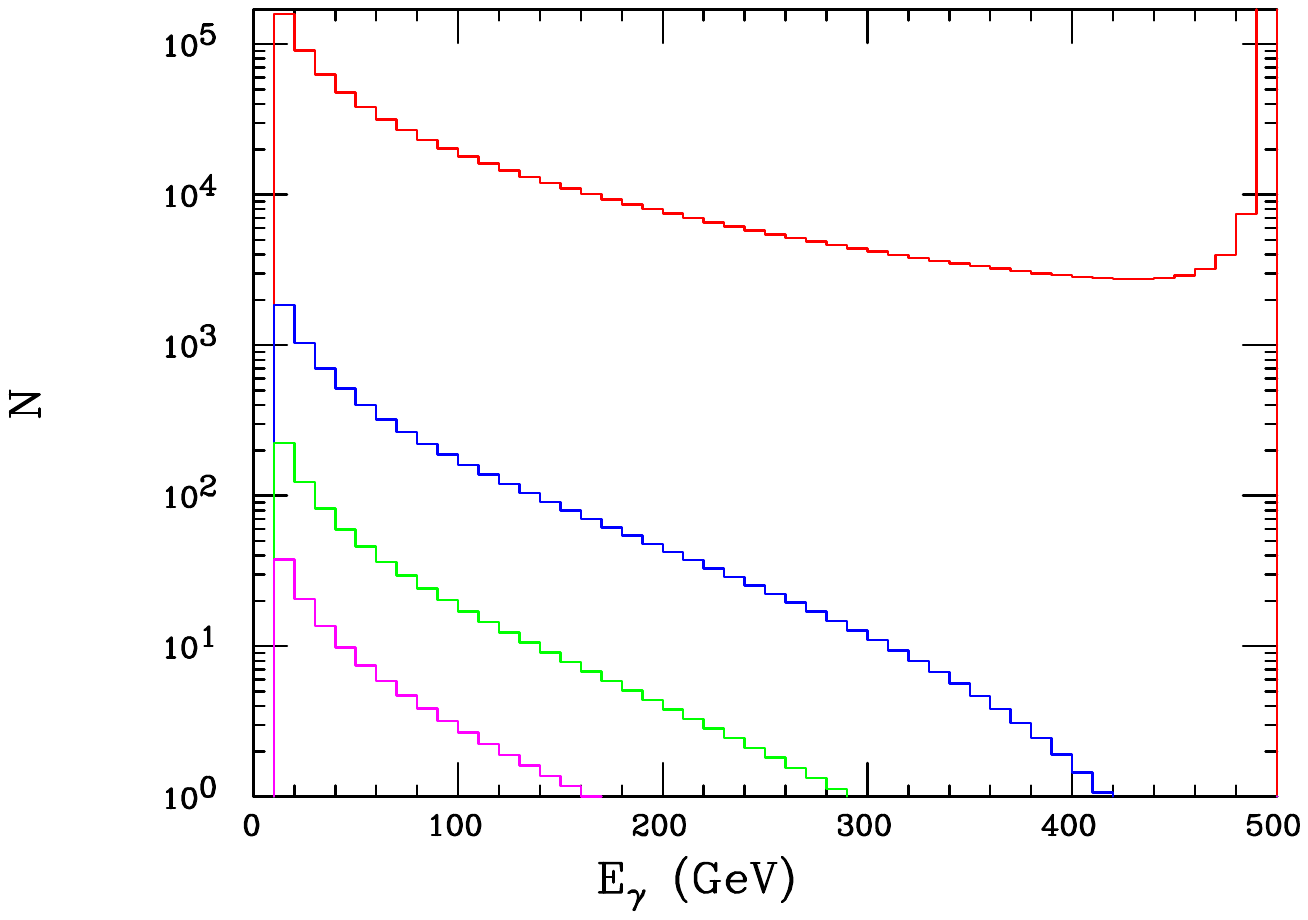}}
\vspace*{-0.8cm}
\centerline{\includegraphics[width=5.0in,angle=0]{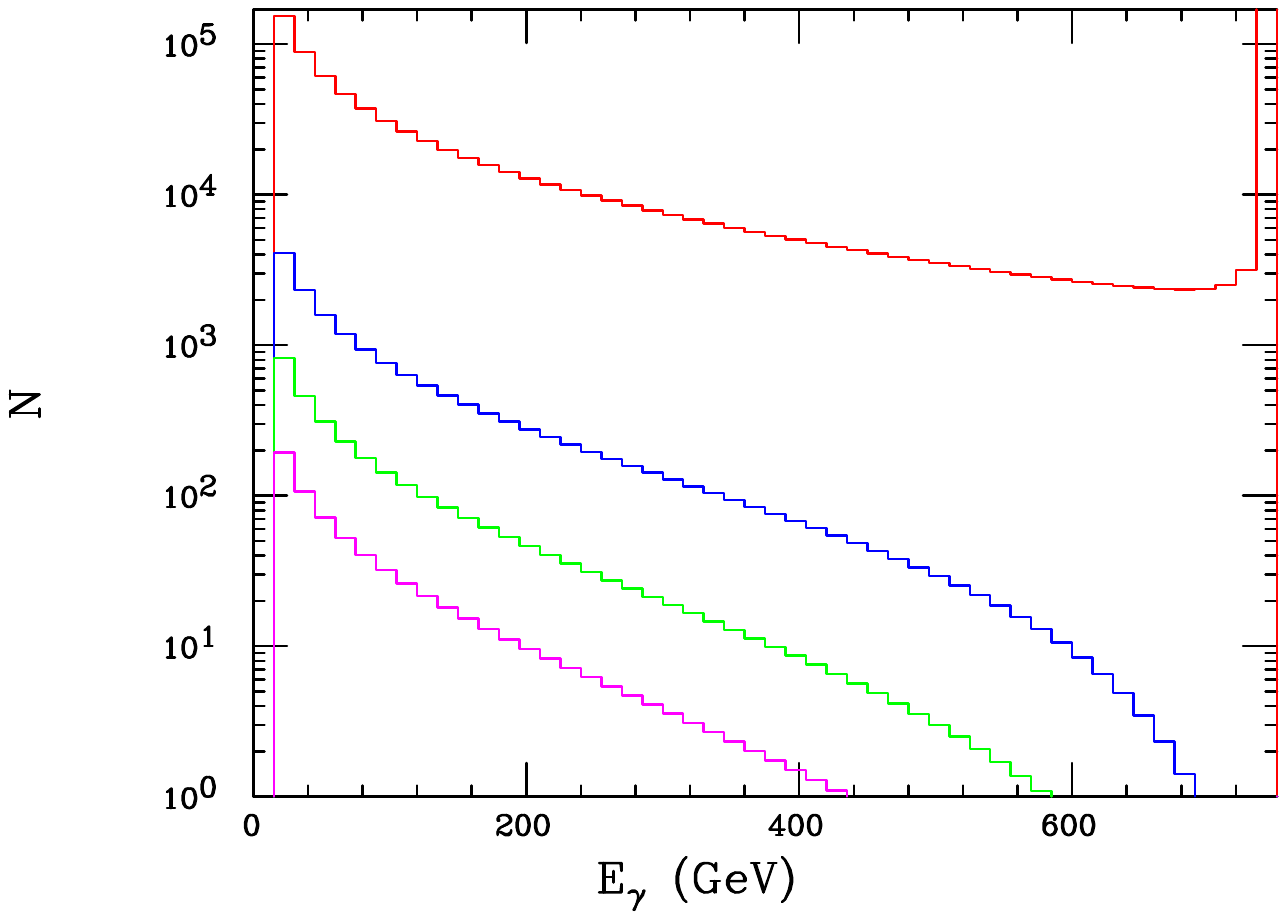}}
\vspace*{-1.3cm}
\caption{(Top) SM background (red) as well as the signal event rates in 10 GeV photon energy bins for the process $e^+e^-\to \gamma +$nothing arising here from dark $Z_i$ and dark Higgs 
production as described in the text assuming $\sqrt s=1$ TeV, an integrated luminosity, $L$, of 1 ab$^{-1}$,  $P_{e^-}=0.8,P_{e^+}=0.3$, $z_0=0.995$, $E_\gamma>10$ GeV and with 
$M_{E_2}=1(1.5,2)$ TeV, corresponding to the blue (green, magenta) histogram for purposes of demonstration. (Bottom) Same as the top panel but now for $\sqrt s =1.5$ TeV in 15 GeV 
photon energy bins and assuming $E_\gamma>15$ GeV.}
\label{fig7}
\end{figure}

In order to make a direct, though rough, comparison of both the signal and background we will require (as done in earlier work) that $E_\gamma >10$ GeV and assume detector coverage down to very 
small angles, \ie, $z_0=0.995$. Further, we will ignore the (small) additional contribution of the heavier PM state, $E_1$, to the total rate with $M_{E_2}\gsim 1$ TeV given the current search 
bound discussed above and the rapid fall off of the cross section with increasing PM mass (hat we will soon observe).  Making these assumptions as well as the preferred polarization choice 
above, we arrive at the results shown in upper panel of Fig.~\ref{fig7} for both the signal and 
background. Here we see that, unfortunately, in the best case scenario, even at lower photon energies the signal falls short of the SM background by over two orders of magnitude when 
$M_{E_2}=1$ TeV and then falls off quite rapidly as either the photon energy or the PM mass increases. The only possible solution is to move closer to the PM pair production threshold and to that 
end we consider going to case of $\sqrt s=1.5$ TeV with the result as shown in the lower panel of this same Figure. Here we see that, while this certainly goes in the right direction by increasing 
$S/B$ by a factor of a few, the signal still remains relatively small. Clearly, unless this SM background is extremely well understood so that per cent level deviation are discernible, it is 
unlikely that the signal will be at all visible.

As we've learned from the discussion above, while the opportunities to directly and/or indirectly probe the physics of the current setup at future colliders is somewhat limited, some interesting 
possibilities remain. Further studies of these and other processes not discussed here are clearly warranted.

\section{Discussion and Conclusion}

The kinetic mixing of the SM photon with a light dark photon associated with a dark gauge group, $U(1)_D$, offers an interesting and flexible window into the possibility of light dark matter and how 
it achieved its observed abundance. This mechanism, however, also requires the existence of a new set of heavy fields, which are either vector-like fermions with respect to the SM gauge symmetries 
and/or complex scalars, that carry both $U(1)_D$ as well as SM couplings, here termed Portal Matter. For a range of parameters, the running of the $U(1)_D$ gauge coupling, especially in the 
case of pseudo-Dirac DM, argues for a possible more UV-complete version of this setup not far from the $\sim10$ TeV scale. There, $U(1)_D$ is embedded into an, at least partially, non-abelian 
group structure, $G_D$, in order to maintain perturbativity by having the opposite sign for its $\beta$-function, with the SM-like $G_D=SU(2)_I \times U(1)_{Y_I}$ being an obvious simple example. 
Although we have examined various aspects of the phenomenology associated with this type of setup in earlier work, it's always been assumed that $G_D$ was broken in a SM-like manner, \ie, directly 
to $U(1)_D$ via a dark Higgs doublet, which then is subsequently broken at a scale $\sim 1$ GeV. Of course this need not be the case. In this paper we have considered the possibility that this $G_D$ 
will be broken instead by a real dark Higgs triplet so that $U(1)_{Y_I}\times U(1)_{T_{3I}}$ remains intact until the low mass scale is reached and thus there are now two light dark gauge bosons. 
Interestingly, if the PM is chosen to transform as an $SU(2)_I$ VLF doublet carrying non-zero SM hypercharge but which is an 
$SU(2)_L$ singlet, then the same vev which breaks $SU(2)_I$ splits the mass degeneracy in this doublet leading to a non-abelian KM between the (still massless) $U(1)_{T_{3I}}$ gauge boson, 
$W_{3I}$, and the familiar hypercharge gauge field, $B$, of the SM. This PM could be `lepton-like', in which case it carries $Q_{em}=-1$ and/or `quark-like', in which case it carries $Q_{em}=2/3$ or 
$-1/3$. Here, we have almost exclusively concentrated on the lepton-like case for simplicity but either case leads to essentially the same calculable value for the KM mixing strength, $\epsilon$, of 
the desired magnitude to satisfy existing experimental and observational constraints. This PM sector is somewhat more minimal than what we've previously encountered, being atypical for several 
reasons and so the resulting phenomenology is also found to be somewhat different.

After shifting the fields to remove the KM and the SM symmetry is broken, the $U(1)_{Y_I}\times U(1)_{T_{3I}}$ remains and requires two Higgs fields to generate the appropriate masses and mix 
the $W_{3I}$ and $B_I$ gauge bosons.  Here these representations were chosen to also perform other tasks for us, \eg, to allow the lighter PM field to decay, it must mix with an analogous SM field 
with the same hypercharge. In the case of lepton-like (quark-like) PM this is just the right-handed charged lepton, $e_R$ (or quark, $u_R,d_R$). The first of these dark Higgs, $H_1$ carrying a 
vev, $v_1$, is thus chosen to be a real $SU(2)_I$ doublet which does this necessary task for us. Also, to generate the Majorana mass term, which together with a Dirac mass for the DM, leads 
to a pseudo-Dirac realization with large mass splitting, $\delta$, hence avoiding multiple direct and indirect search constraints as well as those arising from the CMB, a second dark Higgs, $H_2$, 
is required. Since our DM fermion lies in an $SU(2)_I$ doublet, the required dark Higgs in this case, $H_2$, is correspondingly an isotriplet with $Y_I/2=1$ whose $T_{3I}=1$ member obtains the 
vev, $v_2$. These two vevs, which are assumed to be of comparable value, are then sufficient to completely break the remaining gauge symmetries leaving us with the two light dark gauge 
bosons, $Z_{1,2}$, having somewhat complex coupling structures to dark sector fields. However, they only interact with SM fields via KM and thus their couplings only depend only upon $\epsilon$ 
and the mixing angle, $\phi$, needed to diagonalize the $W_{3I}-B_I$ mass-squared matrix. Demanding that the corresponding mass-squared parameters appearing in this matrix be real was 
found to lead to an additional constraint on the ratio of the $SU(2)_I$ and $U(1)_{Y_I}$ gauge couplings, $g_{Y_I}/g_I$, which leads to some important phenomenological implications. 

The fact that both $Z_i$ will now contribute to the DM relic density calculation (which happens via co-annihilation) was found to open up newly favored regions of model parameter space 
for the ratio of the DM to $Z_1$ masses due to the (mostly) constructive interference between the two exchanges. This region was previously found to require large and so disfavored values of 
$g_I\epsilon$ in order to overcome the relative suppression associated with the $\chi_2$ thermal distribution. Now it is found that this suppression can be overcome without such large couplings 
when the masses of both $Z_i$'s are not too dissimilar, $\lambda_R^2=M_2^2/M_1^2 \lsim 4-5$, with this constructive interference naturally occurring in this parameter regime due to the 
consistency constraints within the setup itself. However, the size of this newly allowed region is also found to depend to some extent upon the assumed values of $\chi_2-\chi_1$ DM relative mass 
spitting, $ \delta$, with smaller values remaining preferred. These same interference effects were also shown to lead to O(1) or larger modifications to the anticipated $\chi_2$ decay lifetime.

The new particles in this model were found to be somewhat harder to access than in the SM-like setup for several reasons, the most important being that the PM in the present case does not 
share an $SU(2)_I$ representation with any of the SM fermions. Thus, \eg, the heavy $W_I$ can no longer be pair produced via the exchange of a heavy $Z_I$ gauge boson (which is absent in 
the current setup) that couples with electroweak strength ($\sim g_I$) to the SM fermions, nor can the $W_I$ be made via associated production together with PM as it doesn't couple to the SM 
fermions at tree level.  The $W_I$ can be pair produced via KM suppressed couplings but it was shown that these cross sections are quite small resulting in very poor search reaches. 
The PM itself can, of course, be pair produced via SM interactions which, in the lepton-like case concentrated on here, occurs as usual via $s$-channel v$\gamma$ and 
$Z$ exchanges and will lead to the slepton-like, opposites-sign dilepton + MET signature at colliders. Existing limits from the LHC are rather poor for these states ($\sim 0.9$ TeV) due to their rather 
small production cross sections (as they are SM isosinglets) and the correspondingly large $W$-pair backgrounds. More serious constraints are, of course, obtainable in the case that the PM is 
quark-like as then it is produced via the strong interactions but in either case, future hadron colliders will be able to significantly extend the search reaches for these states. In the case of the scalar 
sector, the light dark Higgs fields at the $\lsim 1$ GeV scale are rather constrained from mixing with H(125) as they easily lead to a significant branching fraction for decay into the invisible final state.  
The heavier dark Higgs that remains from the real $SU(2)_I$ triplet responsible for the breaking of $SU(2)_I$, $\Sigma^0$, however, is allowed to have a somewhat larger mixing with the SM Higgs 
as the resulting effects are somewhat more subtle, reducing the H(125) width while leaving branching fractions unaltered. This was shown to lead to the possibility that this still small mixing may still be 
sufficiently large to allow for substantial $\Sigma^0$ production at the LHC (for $\lsim 1$ TeV masses) as well as at future colliders by, \eg, $gg$-fusion via the conventional top quark loop while still 
avoiding unitarity and other experimental constraints on H(125) properties. Rare decays of the H(125) into the dark light gauge bosons plus a photon was also shown, however, to be highly suppressed. 
Finally, indirect signatures, \eg, $e^+e^- \to \gamma +$`nothing' at future TeV-scale lepton colliders with $e^\pm$ beam polarization, 
as can arise from the production of the light dark gauge and Higgs bosons in the current setup, were explored but were unfortunately found to fall short of discovery unless the rather large SM 
background were known to a rather high precision. Other mechanisms to obtain information about this sector of the model will require further investigation. 

PM models generally require the existence of extended gauge and Higgs sectors which can lead to many additional new fields and other links between the SM and the dark sector beyond just the 
required KM. Hopefully signals of the dark sector will soon be discovered.

\section*{Acknowledgements}
The author would like to particularly thank J.L. Hewett for a wide range of valuable discussions and the Brookhaven National Laboratory Theory Group for its great hospitality during several visits. This 
work was supported by the Department of Energy, Contract DE-AC02-76SF00515.




\begin{thebibliography}{99}

\bibitem{Carr:2020xqk}
For a recent review, see 
B.~Carr and F.~Kuhnel,
Ann. Rev. Nucl. Part. Sci. \textbf{70}, 355-394 (2020)
doi:10.1146/annurev-nucl-050520-125911
[arXiv:2006.02838 [astro-ph.CO]].


\bibitem{Planck:2018vyg}
N.~Aghanim \textit{et al.} [Planck],
Astron. Astrophys. \textbf{641}, A6 (2020)
[erratum: Astron. Astrophys. \textbf{652}, C4 (2021)]
[arXiv:1807.06209 [astro-ph.CO]].

\bibitem{Kawasaki:2013ae} 
  M.~Kawasaki and K.~Nakayama,
  Ann.\ Rev.\ Nucl.\ Part.\ Sci.\  {\bf 63}, 69 (2013)
  [arXiv:1301.1123 [hep-ph]].
 
\bibitem{Graham:2015ouw} 
  P.~W.~Graham, I.~G.~Irastorza, S.~K.~Lamoreaux, A.~Lindner and K.~A.~van Bibber,
  Ann.\ Rev.\ Nucl.\ Part.\ Sci.\  {\bf 65}, 485 (2015)
  [arXiv:1602.00039 [hep-ex]].

\bibitem{Irastorza:2018dyq}
I.~G.~Irastorza and J.~Redondo,
Prog. Part. Nucl. Phys. \textbf{102}, 89-159 (2018)
[arXiv:1801.08127 [hep-ph]].

\bibitem{Arcadi:2017kky} 
 G.~Arcadi, M.~Dutra, P.~Ghosh, M.~Lindner, Y.~Mambrini, M.~Pierre, S.~Profumo and F.~S.~Queiroz, 
Eur. Phys. J. C \textbf{78}, no.3, 203 (2018)
[arXiv:1703.07364 [hep-ph]].
  
\bibitem{Roszkowski:2017nbc}
L.~Roszkowski, E.~M.~Sessolo and S.~Trojanowski,
Rept. Prog. Phys. \textbf{81}, no.6, 066201 (2018)
[arXiv:1707.06277 [hep-ph]].

\bibitem{Arcadi:2024ukq}
G.~Arcadi, D.~Cabo-Almeida, M.~Dutra, P.~Ghosh, M.~Lindner, Y.~Mambrini, J.~P.~Neto, M.~Pierre, S.~Profumo and F.~S.~Queiroz,
[arXiv:2403.15860 [hep-ph]].
 
\bibitem{LHC}
  K. Pachal,  ``Dark Matter Searches at ATLAS and CMS'', given at the $8^{th}$ {\it {Edition of the Large Hadron Collider Physics Conference}}, 25-30 May, 2020.
 
\bibitem{Aprile:2018dbl}
E.~Aprile \textit{et al.} [XENON],
Phys. Rev. Lett. \textbf{121}, no.11, 111302 (2018)
[arXiv:1805.12562 [astro-ph.CO]]. 

\bibitem{Fermi-LAT:2016uux}
A.~Albert \textit{et al.} [Fermi-LAT and DES],
Astrophys. J. \textbf{834}, no.2, 110 (2017)
[arXiv:1611.03184 [astro-ph.HE]].

\bibitem{Amole:2019fdf}
C.~Amole \textit{et al.} [PICO],
Phys. Rev. D \textbf{100}, no.2, 022001 (2019)
[arXiv:1902.04031 [astro-ph.CO]].

\bibitem{LZ:2022ufs}
J.~Aalbers \textit{et al.} [LZ],
[arXiv:2207.03764 [hep-ex]].

\bibitem{PandaX:2024qfu}
Z.~Bo \textit{et al.} [PandaX],
[arXiv:2408.00664 [hep-ex]].

\bibitem{SuperCDMS:2024yiv}
M.~F.~Albakry \textit{et al.} [SuperCDMS],
[arXiv:2407.08085 [hep-ex]].

\bibitem{Aprile:2024xqi}
E.~Aprile, J.~Aalbers, K.~Abe, S.~Ahmed Maouloud, L.~Althueser, B.~Andrieu, E.~Angelino, D.~A.~Martin, F.~Arneodo and L.~Baudis, \textit{et al.}
[arXiv:2411.15289 [hep-ex]].


\bibitem{Alexander:2016aln} 
  J.~Alexander {\it et al.},
  arXiv:1608.08632 [hep-ph].

\bibitem{Battaglieri:2017aum} 
  M.~Battaglieri {\it et al.},
  arXiv:1707.04591 [hep-ph].
  
\bibitem{Bertone:2018krk}
G.~Bertone and T.~Tait, M.P.,
Nature \textbf{562}, no.7725, 51-56 (2018)
[arXiv:1810.01668 [astro-ph.CO]].

\bibitem{Cooley:2022ufh}
J.~Cooley, T.~Lin, W.~H.~Lippincott, T.~R.~Slatyer, T.~T.~Yu, D.~S.~Akerib, T.~Aramaki, D.~Baxter, T.~Bringmann and R.~Bunker, \textit{et al.}
[arXiv:2209.07426 [hep-ph]].

\bibitem{Boveia:2022syt}
A.~Boveia, T.~Y.~Chen, C.~Doglioni, A.~Drlica-Wagner, S.~Gori, W.~H.~Lippincott, M.~E.~Monzani, C.~Prescod-Weinstein, B.~Shakya and T.~R.~Slatyer, \textit{et al.}
[arXiv:2210.01770 [hep-ph]].


\bibitem{Schuster:2021mlr}
P.~Schuster, N.~Toro and K.~Zhou,
Phys. Rev. D \textbf{105}, no.3, 035036 (2022)
[arXiv:2112.02104 [hep-ph]].

\bibitem{Cirelli:2024ssz}
M.~Cirelli, A.~Strumia and J.~Zupan,
[arXiv:2406.01705 [hep-ph]].

\bibitem{Lanfranchi:2020crw}
G.~Lanfranchi, M.~Pospelov and P.~Schuster,
Ann. Rev. Nucl. Part. Sci. \textbf{71}, 279-313 (2021)
doi:10.1146/annurev-nucl-102419-055056
[arXiv:2011.02157 [hep-ph]].

\bibitem{Steigman:2012nb}
G.~Steigman, B.~Dasgupta and J.~F.~Beacom,
Phys. Rev. D \textbf{86}, 023506 (2012)
[arXiv:1204.3622 [hep-ph]].

\bibitem{Steigman:2015hda} 
  G.~Steigman,
  Phys.\ Rev.\ D {\bf 91}, no. 8, 083538 (2015)
  [arXiv:1502.01884 [astro-ph.CO]].
  
\bibitem{Saikawa:2020swg}
K.~Saikawa and S.~Shirai,
[arXiv:2005.03544 [hep-ph]].

\bibitem{Lee:1977ua}
B.~W.~Lee and S.~Weinberg,
Phys. Rev. Lett. \textbf{39}, 165-168 (1977).

\bibitem{Kolb:1985nn}
E.~W.~Kolb and K.~A.~Olive,
Phys. Rev. D \textbf{33}, 1202 (1986)
[erratum: Phys. Rev. D \textbf{34}, 2531 (1986)]. 




\bibitem{KM}
  B.~Holdom,
  Phys.\ Lett.\  {\bf 166B}, 196 (1986) and
  Phys.\ Lett.\ B {\bf 178}, 65 (1986); 
  K.~R.~Dienes, C.~F.~Kolda and J.~March-Russell,
  Nucl.\ Phys.\ B {\bf 492}, 104 (1997)
  [hep-ph/9610479];
  F.~Del Aguila,
  Acta Phys.\ Polon.\ B {\bf 25}, 1317 (1994)
  [hep-ph/9404323];
  K.~S.~Babu, C.~F.~Kolda and J.~March-Russell,
  Phys.\ Rev.\ D {\bf 54}, 4635 (1996)
  [hep-ph/9603212];
  T.~G.~Rizzo,
  Phys.\ Rev.\ D {\bf 59}, 015020 (1998)
  [hep-ph/9806397].

\bibitem{vectorportal} 
 There has been a huge amount of historical work on this subject; see, for example, 
  D.~Feldman, B.~Kors and P.~Nath,
  Phys.\ Rev.\ D {\bf 75}, 023503 (2007)
  [hep-ph/0610133];
  D.~Feldman, Z.~Liu and P.~Nath,
  Phys.\ Rev.\ D {\bf 75}, 115001 (2007)
  [hep-ph/0702123 [HEP-PH]].;
  M.~Pospelov, A.~Ritz and M.~B.~Voloshin,
  Phys.\ Lett.\ B {\bf 662}, 53 (2008)
  [arXiv:0711.4866 [hep-ph]];
  M.~Pospelov,
  Phys.\ Rev.\ D {\bf 80}, 095002 (2009)
  [arXiv:0811.1030 [hep-ph]]; 
  H.~Davoudiasl, H.~S.~Lee and W.~J.~Marciano,
  Phys.\ Rev.\ Lett.\  {\bf 109}, 031802 (2012)
  [arXiv:1205.2709 [hep-ph]] and 
  Phys.\ Rev.\ D {\bf 85}, 115019 (2012)
  [arXiv:1203.2947 [hep-ph]];
  R.~Essig {\it et al.},
  arXiv:1311.0029 [hep-ph];
  E.~Izaguirre, G.~Krnjaic, P.~Schuster and N.~Toro,
  Phys.\ Rev.\ Lett.\  {\bf 115}, no. 25, 251301 (2015)
  [arXiv:1505.00011 [hep-ph]];
  M.~Khlopov,
  Int.\ J.\ Mod.\ Phys.\ A {\bf 28}, 1330042 (2013)
  [arXiv:1311.2468 [astro-ph.CO]];
 For a general overview and introduction to this framework, see  
  D.~Curtin, R.~Essig, S.~Gori and J.~Shelton,
  JHEP {\bf 1502}, 157 (2015)
  [arXiv:1412.0018 [hep-ph]].

\bibitem{Gherghetta:2019coi}
T.~Gherghetta, J.~Kersten, K.~Olive and M.~Pospelov,
Phys. Rev. D \textbf{100}, no.9, 095001 (2019)
[arXiv:1909.00696 [hep-ph]].
 
\bibitem{Fabbrichesi:2020wbt}
M.~Fabbrichesi, E.~Gabrielli and G.~Lanfranchi,
[arXiv:2005.01515 [hep-ph]].

\bibitem{Graham:2021ggy}
M.~Graham, C.~Hearty and M.~Williams,
[arXiv:2104.10280 [hep-ph]].

\bibitem{Barducci:2021egn}
D.~Barducci, E.~Bertuzzo, G.~Grilli di Cortona and G.~M.~Salla,
JHEP \textbf{12}, 081 (2021)
[arXiv:2109.04852 [hep-ph]].

\bibitem{Li:2024wqj}
S.~Li, J.~M.~Yang, M.~Zhang and R.~Zhu,
[arXiv:2405.18226 [hep-ph]].

\bibitem{CarcamoHernandez:2023wzf}
For a recent review of vector-like fermions, see 
A.~E.~C\'arcamo Hern\'andez, K.~Kowalska, H.~Lee and D.~Rizzo,
[arXiv:2309.13968 [hep-ph]].

\bibitem{CMS:2024bni}
A.~Hayrapetyan \textit{et al.} [CMS],
[arXiv:2405.17605 [hep-ex]].

\bibitem{Alves:2023ufm}
J.~M.~Alves, G.~C.~Branco, A.~L.~Cherchiglia, C.~C.~Nishi, J.~T.~Penedo, P.~M.~F.~Pereira, M.~N.~Rebelo and J.~I.~Silva-Marcos,
Phys. Rept. \textbf{1057}, 1-69 (2024)
[arXiv:2304.10561 [hep-ph]].

\bibitem{Banerjee:2024zvg}
A.~Banerjee, E.~Bergeaas Kuutmann, V.~Ellajosyula, R.~Enberg, G.~Ferretti and L.~Panizzi,
[arXiv:2406.09193 [hep-ph]].

\bibitem{Guedes:2021oqx}
G.~Guedes and J.~Santiago,
JHEP \textbf{01}, 111 (2022)
[arXiv:2107.03429 [hep-ph]].

\bibitem{Adhikary:2024esf}
A.~Adhikary, M.~Olechowski, J.~Rosiek and M.~Ryczkowski,
[arXiv:2406.16050 [hep-ph]].

\bibitem{Benbrik:2024fku}
R.~Benbrik, M.~Boukidi, M.~Ech-chaouy, S.~Moretti, K.~Salime and Q.~S.~Yan,
[arXiv:2412.01761 [hep-ph]].

\bibitem{Albergaria:2024pji}
F.~Albergaria, J.~F.~Bastos, B.~Belfatto, G.~C.~Branco, J.~T.~Penedo, A.~Rodr\'\i{}guez-S\'anchez and J.~I.~Silva-Marcos,
[arXiv:2412.21201 [hep-ph]].


\bibitem{Rizzo:2018vlb}
T.~G.~Rizzo,
Phys. Rev. D \textbf{99}, no.11, 115024 (2019)
[arXiv:1810.07531 [hep-ph]].

\bibitem{Rueter:2019wdf}
T.~D.~Rueter and T.~G.~Rizzo,
Phys. Rev. D \textbf{101}, no.1, 015014 (2020)
[arXiv:1909.09160 [hep-ph]].

\bibitem{Kim:2019oyh}
J.~H.~Kim, S.~D.~Lane, H.~S.~Lee, I.~M.~Lewis and M.~Sullivan,
Phys. Rev. D \textbf{101}, no.3, 035041 (2020)
[arXiv:1904.05893 [hep-ph]].

\bibitem{Rueter:2020qhf}
T.~D.~Rueter and T.~G.~Rizzo,
[arXiv:2011.03529 [hep-ph]].

\bibitem{Wojcik:2020wgm}
G.~N.~Wojcik and T.~G.~Rizzo,
Phys. Rev. D \textbf{105}, no.1, 015032 (2022)
[arXiv:2012.05406 [hep-ph]].

\bibitem{Rizzo:2021lob}
T.~G.~Rizzo,
JHEP \textbf{11}, 035 (2021)
[arXiv:2106.11150 [hep-ph]].

\bibitem{Rizzo:2022qan}
T.~G.~Rizzo,
[arXiv:2202.02222 [hep-ph]].

\bibitem{Wojcik:2022rtk}
G.~N.~Wojcik,
[arXiv:2205.11545 [hep-ph]].

\bibitem{Rizzo:2022jti}
T.~G.~Rizzo,
Phys. Rev. D \textbf{106}, no.3, 035024 (2022)
[arXiv:2206.09814 [hep-ph]].

\bibitem{Rizzo:2022lpm}
T.~G.~Rizzo,
Phys. Rev. D \textbf{106}, no.9, 095024 (2022)
[arXiv:2209.00688 [hep-ph]].

\bibitem{Carvunis:2022yur}
A.~Carvunis, N.~McGinnis and D.~E.~Morrissey,
[arXiv:2209.14305 [hep-ph]].

\bibitem{Verma:2022nyd}
S.~Verma, S.~Biswas, A.~Chatterjee and J.~Ganguly,
[arXiv:2209.13888 [hep-ph]].

\bibitem{Rizzo:2023qbj}
T.~G.~Rizzo,
Phys. Rev. D \textbf{107}, no.9, 095014 (2023)
[arXiv:2302.12698 [hep-ph]].

\bibitem{Wojcik:2022woa}
G.~N.~Wojcik, L.~L.~Everett, S.~T.~Eu and R.~Ximenes,
Phys. Lett. B \textbf{841}, 137931 (2023)
[arXiv:2211.09918 [hep-ph]].

\bibitem{Wojcik:2023ggt}
G.~N.~Wojcik, L.~L.~Everett, S.~T.~Eu and R.~Ximenes,
Phys. Rev. D \textbf{108}, no.5, 055033 (2023)
[arXiv:2303.12983 [hep-ph]].


\bibitem{Rizzo:2023kvy}
T.~G.~Rizzo,
Phys. Rev. D \textbf{108}, no.5, 055021 (2023)
[arXiv:2307.08508 [hep-ph]].

\bibitem{Rizzo:2023djp}
T.~G.~Rizzo,
Phys. Rev. D \textbf{109}, no.5, 055039 (2024)
[arXiv:2312.00226 [hep-ph]].

\bibitem{Rizzo:2024bhn}
T.~G.~Rizzo,
Phys. Rev. D \textbf{110}, no.7, 075037 (2024)
[arXiv:2408.01296 [hep-ph]].

\bibitem{Ardu:2024bxg}
M.~Ardu, M.~H.~Rahat, N.~Valori and O.~Vives,
[arXiv:2407.21100 [hep-ph]].

\bibitem{Rizzo:2024kzu}
T.~G.~Rizzo,
Phys. Rev. D \textbf{111}, no.7, 075018 (2025)
[arXiv:2412.17174 [hep-ph]].

\bibitem{Slatyer:2015jla}
T.~R.~Slatyer,
Phys. Rev. D \textbf{93}, no.2, 023527 (2016)
[arXiv:1506.03811 [hep-ph]].

\bibitem{Liu:2016cnk} 
  H.~Liu, T.~R.~Slatyer and J.~Zavala,
  Phys.\ Rev.\ D {\bf 94}, no. 6, 063507 (2016)
  [arXiv:1604.02457 [astro-ph.CO]].

\bibitem{Leane:2018kjk}
R.~K.~Leane, T.~R.~Slatyer, J.~F.~Beacom and K.~C.~Ng,
Phys. \ Rev. \ D \textbf{98}, no.2, 023016 (2018)
[arXiv:1805.10305 [hep-ph]].

\bibitem{Wang:2025tdx}
Y.~N.~Wang, X.~C.~Duan, T.~P.~Tang, Z.~Wang and Y.~L.~S.~Tsai,
[arXiv:2502.18263 [hep-ph]].

\bibitem{Koechler:2023ual}
J.~Koechler,
[arXiv:2309.10043 [hep-ph]].

\bibitem{DelaTorreLuque:2023cef}
P.~De la Torre Luque, S.~Balaji and J.~Silk,
[arXiv:2312.04907 [hep-ph]].

\bibitem{Wang:2025jhy}
G.~Wang, B.~Y.~Su, L.~Zu and L.~Feng,
[arXiv:2503.22148 [astro-ph.HE]].

\bibitem{Belanger:2024bro}
G.~B\'elanger, S.~Chakraborti, Y.~G\'enolini and P.~Salati,
[arXiv:2401.02513 [hep-ph]].

\bibitem{Brahma:2023psr}
N.~Brahma, S.~Heeba and K.~Schutz,
Phys. Rev. D \textbf{109}, no.3, 035006 (2024)
[arXiv:2308.01960 [hep-ph]].

\bibitem{Balan:2024cmq}
S.~Balan, C.~Bal\'azs, T.~Bringmann, C.~Cappiello, R.~Catena, T.~Emken, T.~E.~Gonzalo, T.~R.~Gray, W.~Handley and Q.~Huynh, \textit{et al.}
[arXiv:2405.17548 [hep-ph]].

\bibitem{Garcia:2024uwf}
See, for example, G.~D.~V.~Garcia, F.~Kahlhoefer, M.~Ovchynnikov and T.~Schwetz,
[arXiv:2405.08081 [hep-ph]]. 
 
\bibitem{Mohlabeng:2024itu}
G.~Mohlabeng, A.~Mondol and T.~M.~P.~Tait,
[arXiv:2405.08881 [hep-ph]].


\bibitem{Davoudiasl:2015hxa}
H.~Davoudiasl and W.~J.~Marciano,
Phys. Rev. D \textbf{92}, no.3, 035008 (2015)
[arXiv:1502.07383 [hep-ph]].

\bibitem{Reilly:2023frg}
A.~Reilly and N.~Toro,
[arXiv:2308.01347 [hep-ph]].

\bibitem{Bauer:2022nwt}
For related work on the possibilities of KM and DM physics employing this same gauge group, see 
M.~Bauer and P.~Foldenauer,
Phys. Rev. Lett. \textbf{129}, no.17, 171801 (2022)
[arXiv:2207.00023 [hep-ph]].

\bibitem{Hewett:1988xc}
J.~L.~Hewett and T.~G.~Rizzo,
Phys. Rept. \textbf{183}, 193 (1989).


\bibitem{ATLAS:2023tkt}
G.~Aad \textit{et al.} [ATLAS],
Phys. Lett. B \textbf{842}, 137963 (2023)
[arXiv:2301.10731 [hep-ex]].

\bibitem{CMS:2023sdw}
A.~Tumasyan \textit{et al.} [CMS],
Eur. Phys. J. C \textbf{83}, no.10, 933 (2023)
[arXiv:2303.01214 [hep-ex]].

\bibitem{ATLAS:2024zxk}
G.~Aad \textit{et al.} [ATLAS],
[arXiv:2407.09168 [hep-ex]].

\bibitem{CMS:2024jyb}
A.~Hayrapetyan \textit{et al.} [CMS],
[arXiv:2407.20425 [hep-ex]].

\bibitem{Rizzo:2020jsm}
See, for example, T.~G.~Rizzo,
JHEP \textbf{01}, 079 (2021)
[arXiv:2006.08502 [hep-ph]].

\bibitem{Giovanetti:2021izc}
C.~Giovanetti, M.~Lisanti, H.~Liu and J.~T.~Ruderman,
Phys. Rev. Lett. \textbf{129}, no.2, 021302 (2022)
[arXiv:2109.03246 [hep-ph]].

\bibitem{Chu:2022xuh}
X.~Chu, J.~L.~Kuo and J.~Pradler,
Phys. Rev. D \textbf{106}, no.5, 055022 (2022)
[arXiv:2205.05714 [hep-ph]].

\bibitem{Sabti:2021reh}
N.~Sabti, J.~Alvey, M.~Escudero, M.~Fairbairn and D.~Blas,
JCAP \textbf{08}, A01 (2021)
[arXiv:2107.11232 [hep-ph]].

\bibitem{Sabti:2019mhn}
N.~Sabti, J.~Alvey, M.~Escudero, M.~Fairbairn and D.~Blas,
JCAP \textbf{01}, 004 (2020)
[arXiv:1910.01649 [hep-ph]].

\bibitem{Rizzo:2012rb}
T.~G.~Rizzo,
Phys. Rev. D \textbf{86}, 055024 (2012)
[arXiv:1206.7055 [hep-ph]].


\bibitem{Rizzo:2021pxo}
T.~G.~Rizzo,
JHEP \textbf{04}, 248 (2021)
[arXiv:2102.03647 [hep-ph]].


\bibitem{ATLAS:2019erb}
G.~Aad \textit{et al.} [ATLAS],
Phys. Lett. B \textbf{796}, 68-87 (2019)
[arXiv:1903.06248 [hep-ex]].

\bibitem{CMS:2021ctt}
A.~M.~Sirunyan \textit{et al.} [CMS],
JHEP \textbf{07}, 208 (2021)
[arXiv:2103.02708 [hep-ex]].

\bibitem{Helsens:2019bfw}
C.~Helsens, D.~Jamin, M.~L.~Mangano, T.~G.~Rizzo and M.~Selvaggi,
Eur. Phys. J. C \textbf{79}, 569 (2019)
[arXiv:1902.11217 [hep-ph]].

\bibitem{ParticleDataGroup:2024cfk}
S.~Navas \textit{et al.} [Particle Data Group],
Phys. Rev. D \textbf{110}, no.3, 030001 (2024)

\bibitem{Tackmann:2023wfs}
K.~Tackmann,
LHEP \textbf{2023}, 440 (2023).

\bibitem{Moriond25}
For a recent update on the Higgs boson couplings, see the talk given by F. Monti on behalf of the CMS and ATLAS Collaborations at the  
``$59^{th}$ Rencontres de Moriond 2025: QCD and High Energy Interactions'', LaThuile, Italy, 3/30-4/6/25.

\bibitem{Heo:2024cif}
Y.~Heo, D.~W.~Jung and J.~S.~Lee,
Phys. Rev. D \textbf{110}, no.1, 013003 (2024)
[arXiv:2402.02822 [hep-ph]].

\bibitem{Heo:2025vkz}
Y.~Heo, J.~S.~Lee and C.~B.~Park,
[arXiv:2502.02992 [hep-ph]].

\bibitem{Kang:2013zba}
S.~K.~Kang and J.~Park,
JHEP \textbf{04}, 009 (2015)
[arXiv:1306.6713 [hep-ph]].



 \bibitem{GBET}
  M.~S.~Chanowitz and M.~K.~Gaillard,
  Nucl.\ Phys.\ B {\bf 261}, 379 (1985);
  B.~W.~Lee, C.~Quigg and H.~B.~Thacker,
  Phys.\ Rev.\ D {\bf 16}, 1519 (1977);
  J.~M.~Cornwall, D.~N.~Levin and G.~Tiktopoulos,
  Phys.\ Rev.\ D {\bf 10}, 1145 (1974)
  Erratum: [Phys.\ Rev.\ D {\bf 11}, 972 (1975)];
  G.~J.~Gounaris, R.~Kogerler and H.~Neufeld,
  Phys.\ Rev.\ D {\bf 34}, 3257 (1986).


\bibitem{Chu:2018qrm}
X.~Chu, J.~Pradler and L.~Semmelrock,
Phys. Rev. D \textbf{99}, no.1, 015040 (2019)
[arXiv:1811.04095 [hep-ph]].

\bibitem{Choi:2015zka}
S.~Y.~Choi, T.~Han, J.~Kalinowski, K.~Rolbiecki and X.~Wang,
Phys. Rev. D \textbf{92}, no.9, 095006 (2015)
[arXiv:1503.08538 [hep-ph]].


\bibitem{Hirsch:2002uv}
M.~Hirsch, E.~Nardi and D.~Restrepo,
Phys. Rev. D \textbf{67}, 033005 (2003)
[arXiv:hep-ph/0210137 [hep-ph]].

\bibitem{Barranco:2007ej}
J.~Barranco, O.~G.~Miranda, C.~A.~Moura and J.~W.~F.~Valle,
Phys. Rev. D \textbf{77}, 093014 (2008)
[arXiv:0711.0698 [hep-ph]].

\bibitem{Berezhiani:2001rs}
Z.~Berezhiani and A.~Rossi,
Phys. Lett. B \textbf{535}, 207-218 (2002)
[arXiv:hep-ph/0111137 [hep-ph]].

\bibitem{Escrihuela:2019mot}
F.~J.~Escrihuela, L.~J.~Flores and O.~G.~Miranda,
Phys. Lett. B \textbf{802}, 135241 (2020)
[arXiv:1907.12675 [hep-ph]].


\bibitem{ILC:2007oiw}
J.~Brau \textit{et al.} [ILC],
[arXiv:0712.1950 [physics.acc-ph]].

\bibitem{ILCInternationalDevelopmentTeam:2022izu}
A.~Aryshev \textit{et al.} [ILC International Development Team],
[arXiv:2203.07622 [physics.acc-ph]].

\bibitem{Aicheler:2012bya}
M.~Aicheler, P.~Burrows, M.~Draper, T.~Garvey, P.~Lebrun, K.~Peach, N.~Phinney, H.~Schmickler, D.~Schulte and N.~Toge,
CERN-2012-007.

\bibitem{Bai:2021rdg}
M.~Bai, T.~Barklow, R.~Bartoldus, M.~Breidenbach, P.~Grenier, Z.~Huang, M.~Kagan, J.~Lewellen, Z.~Li and T.~W.~Markiewicz, \textit{et al.}
[arXiv:2110.15800 [hep-ex]].

\bibitem{LinearColliderVision:2025hlt}
D.~Atti\'e \textit{et al.} [Linear Collider Vision],
[arXiv:2503.19983 [hep-ex]].




\bibitem{Chae:2012bq}
Y.~J.~Chae and M.~Perelstein,
JHEP \textbf{05}, 138 (2013)
[arXiv:1211.4008 [hep-ph]].

\bibitem{Kalinowski:2021tyr}
J.~Kalinowski, W.~Kotlarski, K.~Mekala, P.~Sopicki and A.~F.~Zarnecki,
Eur. Phys. J. C \textbf{81}, no.10, 955 (2021)
[arXiv:2107.11194 [hep-ph]].

\bibitem{Khalil:2021afa}
S.~Khalil, S.~Moretti, D.~Rojas-Ciofalo and H.~Waltari,
Phys. Rev. D \textbf{104}, no.3, 035008 (2021)
[arXiv:2104.07347 [hep-ph]].

\bibitem{Black:2022qlg}
K.~Black, T.~Bose, Y.~Chen, S.~Dasu, H.~Jia, D.~Pinna, V.~Sharma, N.~Venkatasubramanian and C.~Vuosalo,
[arXiv:2205.10404 [hep-ex]].

\bibitem{Dreiner:2012xm}
H.~Dreiner, M.~Huck, M.~Kr\"amer, D.~Schmeier and J.~Tattersall,
Phys. Rev. D \textbf{87}, no.7, 075015 (2013)
[arXiv:1211.2254 [hep-ph]].






\end{thebibliography}
\end{document}